\documentclass[twocolumn,amsmath,amssymb,nofootinbib,superscriptaddress,longbibliography]{revtex4-1}

\usepackage{graphicx}
\usepackage{amsmath}
\usepackage{amssymb}
\usepackage{xcolor}
\usepackage{hyperref}
\hypersetup{pdfstartview={FitH},pdfpagemode={UseNone}, breaklinks=true,
            colorlinks=true,bookmarks=false,linkcolor=blue, citecolor=blue, urlcolor=blue,
             bookmarksopen=true, pdfnewwindow=true}
\usepackage[all]{hypcap}
\usepackage[english]{babel}
\usepackage{physics}

\begin{document}

\preprint{APS/123-QED}

\title{Generating ${\rm N00N}$-states of surface plasmon-polariton pairs with a nanoparticle}

\author{Nikita A. Olekhno}
 \email{nikita.olekhno@metalab.ifmo.ru}
\affiliation{Department of Physics and Engineering, ITMO University, 49 Kronverksky ave., Saint Petersburg 197101, Russia}
\author{Mihail I. Petrov}
\affiliation{Department of Physics and Engineering, ITMO University, 49 Kronverksky ave., Saint Petersburg 197101, Russia}
\author{Ivan V. Iorsh}
\affiliation{Department of Physics and Engineering, ITMO University, 49 Kronverksky ave., Saint Petersburg 197101, Russia}
\author{Andrey A. Sukhorukov}
\affiliation{ARC Centre of Excellence for Transformative Meta-Optical Systems (TMOS), Research School of Physics, The Australian~National~University, Canberra,~ACT 2601, Australia}
\author{Alexander S. Solntsev}
\affiliation{School of Mathematical and Physical Sciences, University of Technology Sydney, 15 Broadway, Ultimo NSW 2007, Australia}


\date{\today}


\begin{abstract}
We consider a generation of two-particle quantum states in the process of spontaneous parametric down-conversion of light by a dielectric nanoparticle with $\chi^{(2)}$ response. As a particular example, we study the generation of surface plasmon-polariton pairs with a ${\rm GaAs}$ nanoparticle located at the silver-air interface. We show that for certain excitation geometries, ${\rm N00N}$-states of surface plasmon-polariton pairs could be obtained. The effect can be physically interpreted as a result of quantum interference between pairs of induced sources, each emitting either signal or idler plasmon. We then relate the resulting ${\rm N00N}$-pattern to the general symmetry properties of dyadic Green's function of a dipole emitter exciting surface waves. It renders the considered effect as a general way towards a robust generation of ${\rm N00N}$-states of surface waves using spontaneous parametric down-conversion in $\chi^{(2)}$ nanoparticles.
\end{abstract}

\maketitle


\section{Introduction}

Generation of entangled photons and their correlation measurements became one of the central topics in quantum optics addressing both fundamental aspects of quantum mechanics and driving the development of novel solutions in quantum technologies~\cite{2014_Muller, 2014_Versteegh}. By now, the generation of entangled photons has been successfully downscaled from free space experiments to on-chip dielectric waveguides~\cite{2017_Solntsev} enabling integrated quantum photonics applications ranging from quantum information processing ~\cite{2009_OBrien} to spectroscopy~\cite{2018_Solntsev}.

Moreover, the recent experiments have shown that the surface plasmon-polaritons (SPPs) preserve the quantum coherence~\cite{2014_Fakonas, 2016_Dheur, 2017_Dheur} and can transfer the quantum excitations despite the intrinsically strong dissipation (and related fluctuations) and collective nature of the single excitation,  allowing for long-range quantum interference~\cite{2019_DAmico}. Among various two-particle entangled quantum states, ${\rm N00N}$-states attract considerable attention due to their prospective for lowering the quantum noise level below the shot noise limit~\cite{2015_Dowling}. Such states were recently studied in systems ranging from microwave setups based on superconducting qubits \cite{2020_Kannan} to topological photonic crystals \cite{2016_Rechtsman}, ring resonator waveguides \cite{2020_Han}, and plasmonic setups \cite{2018_Chen, 2018_Vest, 2020_Mehta}. Owing to strong spatial localization of SPP modes, they are known to be effectively utilized for optical sensing applications, and implementation of quantum SPP states opens a way for low-noise quantum sensing applications ~\cite{2018_Chen, 2018_Lee, 2015_Fan, 2019_Lawrie}. The current approach to the generation of plasmonic ${\rm N00N}$-states is based on spatial separation of entangled photonic modes, i.e., by using beam splitters and cascaded Mach-Zehnder interferometers, which later are transferred to SPPs.

The most common mechanism of entangled quantum state generation is the second-order nonlinear process of spontaneous parametric down-conversion (SPDC). In the last few years, materials with strong second-order optical nonlinearity such as gallium arsenide (${\rm GaAs}$) and aluminum gallium arsenide (${\rm AlGaAs}$) have attracted significant attention, because they provide a drastic increase of nonlinear optical interaction efficiency on the sub-micron scales~\cite{2016_Camacho_Morales, 2017_Carletti, 2018_Liu, 2020_Saerens}. That resulted in the observation of the entangled photon pairs generation from a single resonant ${\rm AlGaAs}$ nanodisk supporting Mie resonant modes~\cite{2019_Marino}. The generation of entangled photon pairs via SPDC is experimentally demonstrated in various micro- and nanostructured nonlinear materials ranging from thin films \cite{2019_Okoth, 2020_Santiago_Cruz} and metalens arrays \cite{2020_Li} to resonant metasurfaces \cite{2021_Santiago_Cruz}. The motivation for the present study was also strongly supported by the recent advances in directional excitation of SPP modes with dielectric nanoparticles~\cite{2017_Sinev, 2020_Sinev}, which opens perspectives for nonlinear excitation of entangled SPPs.

\begin{figure}[b]
    \centering
    \includegraphics[width=8cm]{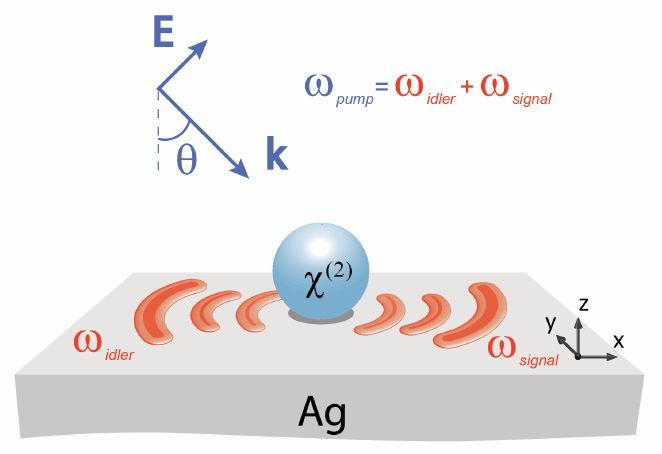}
    \caption{Geometry of the process. A ${\rm GaAs}$ nanoparticle demonstrating second-order nonlinear response is located at the silver-air interface. Excitation of the nanoparticle with a classical pump light in ${\rm TM}$-geometry causes a creation of two surface plasmon-polaritons. The angle between the pump wave vector and the interface normal is $\theta$.}
    \label{fig:Geometry}
\end{figure}

In this article, we consider the generation of ${\rm N00N}$-states through the nonlinear decay of a pump photon at a nanoparticle made of $\chi^{(2)}$ material accompanied by a directional excitation of SPPs at the metal-dielectric interface, Figure~\ref{fig:Geometry}. Following the general Green's function formalism outlined earlier~\cite{2016_Poddubny, 2018_Lenzini, 2018_Poddubny}, we demonstrate that one can obtain ${\rm N00N}$-states of SPP pairs by using a ${\rm GaAs}$ nanoparticle, and that this result remains robust for a wide range of excitation conditions owing to general properties of Green's function of the problem.

The paper is organized as follows. In Section~\ref{sec:Theory}, we present the theoretical framework 
and introduce the model assumptions used in our analysis. Then, in Section~\ref{sec:Results} we analyse the obtained results, including the shape of two-plasmon wave functions and their dependence on the SPDC degeneracy and pump geometry. Section~\ref{sec:Conclusion} contains conclusions and final remarks.

\section{Theoretical Model}
\label{sec:Theory}

{\it General formalism}. We focus on the generation of SPP pairs via the spontaneous parametric down-conversion of pump photons. In the course of this process requiring second-order nonlinearity, one pump photon at the frequency $\omega_{\rm pump}$ is converted into two excitations, idler at the frequency $\omega_{\rm i}$ and signal at the frequency $\omega_{\rm s}$. The energy conservation condition defining the frequencies of down-converted excitations reads as $\hbar\omega_{\rm pump} = \hbar\omega_{\rm i} + \hbar\omega_{\rm s}$. For the sake of simplicity, we start with considering the degenerate SPDC process: $\omega_{\rm i} = \omega_{\rm s} = \omega_{\rm pump}/2$.

Our description is based on the dyadic Green's function formalism developed in Ref.~\cite{2016_Poddubny}. The considered technique involves explicitly field detectors, and thus allows taking into account the damping of propagating waves, which is crucial for the studying SPPs generation. Within this approach, spatial correlations between the generated photons or SPPs can be characterized by  two-particle count rate $W = \frac{2\pi}{\hbar}|T(\mathbf{r}_{\rm i}, \omega_{\rm i}, \mathbf{d}_{\rm i}, \mathbf{r}_{\rm s}, \omega_{\rm s}, \mathbf{d}_{\rm s})|^2\delta(\hbar\omega_{\rm pump} - \hbar\omega_{\rm i} - \hbar\omega_{\rm s})$, which describes the simultaneous detection of two photons or plasmon-polaritons. In the relation above, $\mathbf{r}_{\rm i}$ and $\mathbf{r}_{\rm s}$ are the positions of idler and signal detectors, respectively, $\mathbf{d}_{\rm i}$ and $\mathbf{d}_{\rm s}$ are their dipole moments, and $\omega_{\rm i}$, $\omega_{\rm s}$ are the frequencies of excitations being detected. Delta function represents the energy conservation during the SPDC process. The introduced quantity $W$ allows describing spatial correlations between the idler and signal photons. It is proportional to the square of the two-photon amplitude $T$, which can be expressed as (see Ref.~\cite{2016_Poddubny} and Supplementary Material therein) $T = \int_{V_0} \bra{d^{({\rm i})}}G_{\alpha\beta}(\mathbf{r}_{\rm i}, \mathbf{r}_0, \omega_{\rm i})\Gamma_{\beta\gamma}(\mathbf{r}_0)G_{\gamma\delta}(\mathbf{r}_0, \mathbf{r}_{\rm s}, \omega_{\rm s})\ket{d^{({\rm s})}}d\mathbf{r}_0$. The integration is performed over the entire volume $V_0$ of the nonlinear system generating entangled pairs, $G_{\alpha\beta}(\mathbf{r}, \mathbf{r}', \omega)$ is a dyadic Green's function for the corresponding environment and $\Gamma_{\alpha\beta}(\mathbf{r})$ is the generation matrix. The latter is related to the second-order nonlinear susceptibility tensor $\hat{\chi}^{(2)}$ as
\begin{equation}
    \Gamma_{\alpha\beta}(\mathbf{r}) = \chi_{\alpha\beta\gamma}^{(2)}(\mathbf{r})\mathbf{E}_{\gamma}^{ \rm pump}(\mathbf{r}),
    \label{eq:Gamma}
\end{equation}
with $\mathbf{E}^{\rm pump}(\mathbf{r}){\rm e}^{-i\omega_{\rm pump}t}$ being the electric field of a classical pump. Thus, spatial correlations within the generated pair are sufficiently affected by the form of the susceptibility tensor of generating system.


\begin{figure}[t]
    \centering
    \includegraphics[width=7cm]{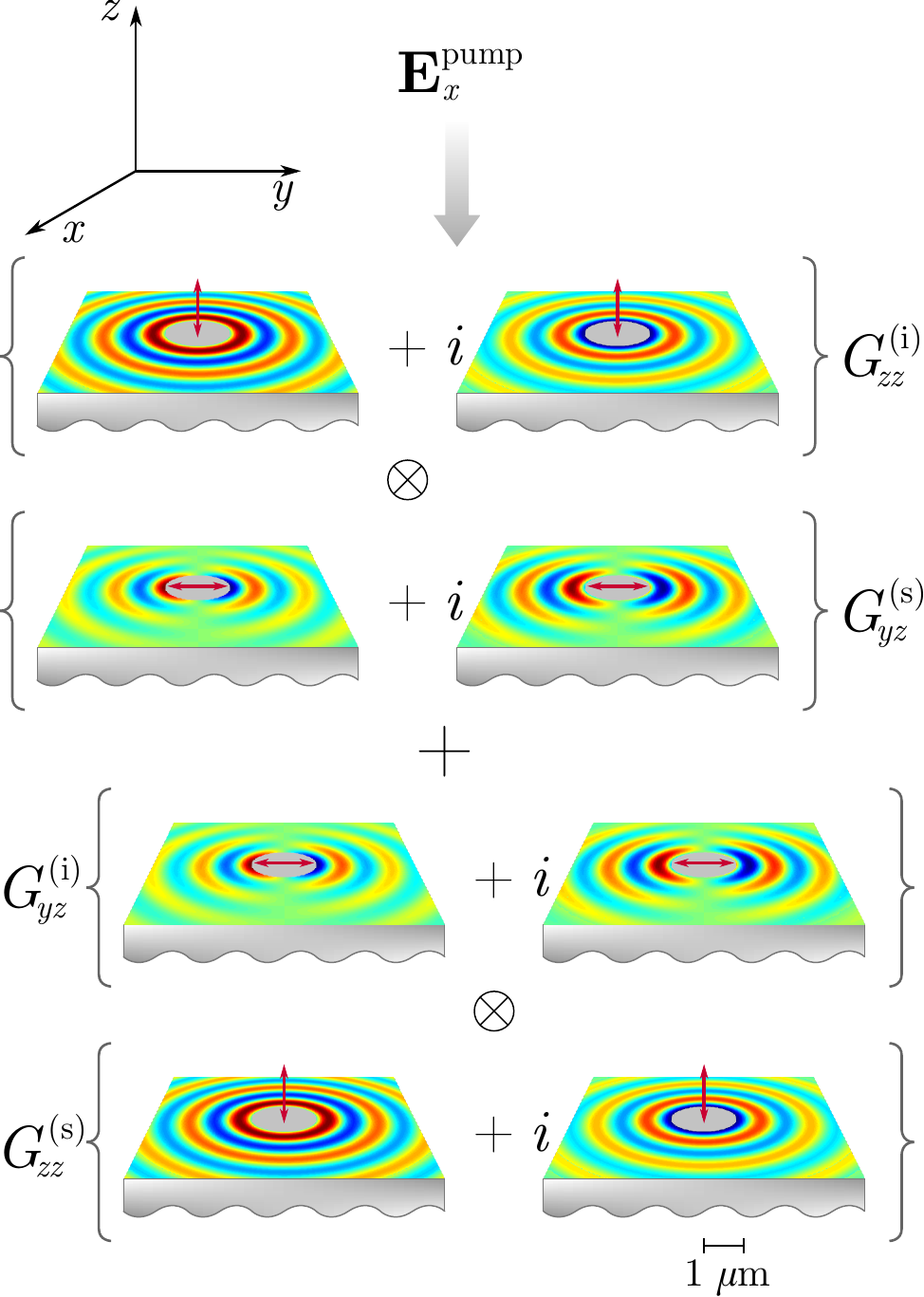}
    \caption{Illustration of the induced dipole decomposition in the case of a  ${\rm GaAs}$ nanoparticle with axis $z||[001]$ irradiated at the normal incidence with pump wave having the wavelength $\lambda_{\rm pump}=750 {\rm nm}$ and linearly polarized along the $x$-axis. The two dipolar sources exciting SPPs at wavelengths $\lambda_{\rm i,s}=2\lambda_{\rm pump}$ are directed along the $z$- and $y$-axis, according to Eq.~(\ref{eq:T_N00N}). Real and imaginary parts of the corresponding dyadic Green's functions $G_{zz}$ and $G_{yz}$ are shown as projections to the interface plane for coordinates $x$,$y$ in the range from $-4$ to $4$ ${\rm \mu m}$. The dipoles, as well as the image planes, are located at the height $z=10~{\rm nm}$.}
    \label{fig:Induced_Dipoles}
\end{figure}

{\it Dipole approximation}. Before we proceed with a consideration of the surface plasmon-polariton quantum states, we first give a more transparent and physical description of the two-photon amplitude. We assume that the size of the nonlinear scatterer is much smaller than the wavelength and, thus, it can be treated in the dipole approximation. Then, the nonlinear tensor will have a simple form $ \chi_{ijk}^{(2)}(\mathbf{r})=\chi_{ijk}^{(2)}V_0\delta(\mathbf{r}-\mathbf{r}_0)$ , where $\mathbf{r}_0$ is the position of the dipole scatterer, and $V_0$ is its volume. The two-particle amplitude $T$ takes the form
\begin{equation}
    T = V_0\bra{d^{({\rm i})}}G_{\alpha\beta}(\mathbf{r}_{\rm i}, \mathbf{r}_0, \omega_{\rm i})\Gamma_{\beta\gamma}(\mathbf{r}_0)G_{\gamma\delta}(\mathbf{r}_0, \mathbf{r}_{\rm s}, \omega_{\rm s})\ket{d^{({\rm s})}}.
    \label{eq:T}
\end{equation}
Thus, spatial correlations between the  down-converted SPP pair are defined by the matrix elements of the convolution of three tensors, two characterizing the Green's functions at the given detection positions $\mathbf{r}_{\rm i}$ and $\mathbf{r}_{\rm s}$ and the central one describing the nonlinear generation process. Next,  by decomposing the generation matrix $\hat \Gamma$ into dyads:
\begin{equation}
    \hat{\Gamma} = \sum_{\alpha,\beta = \{x,y,z\}}\Gamma_{\alpha\beta}\ket{\mathbf{e}_{\alpha}}\bra{\mathbf{e}_\beta},
\end{equation}
and decomposing the dipole moments of detectors within the same Cartesian basis vectors $\mathbf{e}_{x,y,z}$, $\bra{d^{({\rm i})}} = \sum_{\alpha}d_{\alpha}^{(i)*}\bra{\mathbf{e}_{\alpha}}$, $\ket{d^{({\rm i})}} = \sum_{\beta}d_{\beta}^{({\rm i})}\ket{\mathbf{e}_{\beta}}$ with indices $\alpha$, $\beta$ taking values $x$, $y$ and $z$, Eq.~(\ref{eq:T}) renders as a sum of pairwise products involving dyadic Green's function matrix elements having the form $\bra{\mathbf{e}_{\alpha}}\hat{G}(\mathbf{r}_{\rm i},\mathbf{r}_0,\omega_{\rm i})\ket{\mathbf{e}_{\gamma}}\bra{\mathbf{e}_{\delta}}\hat{G}(\mathbf{r}_{\rm s},\mathbf{r}_0,\omega_{\rm s})\ket{\mathbf{e}_{\beta}} = G_{\alpha\gamma}^{({\rm i})}G_{\beta\delta}^{({\rm s})}$. The two-photon amplitude can be decomposed as follows: 
\begin{equation}
    T=V_0\sum_{\alpha\beta\gamma\delta}d_{\alpha}^{({\rm i})*} G_{\alpha\beta}^{({\rm i})}\Gamma_{\beta\gamma}G_{\gamma\delta}^{({\rm s})}d_{\delta}^{({\rm s})}
\end{equation}
Thereby, the spatial correlations between the generated photon (plasmon) pairs can be represented as a result of {\it interference} of electric fields produced by pairs of dipole emitters each emitting one excitation at half the pump frequency, Figure~\ref{fig:Induced_Dipoles}. The directions  of {\it induced dipoles}  are totally defined by the structure of the nonlinear tensor $\hat\chi^{(2)}$, direction and polarization of the pump wave. The detecting dipoles $d^{({\rm i})}$ and $d^{({\rm s})}$ define the particular polarizations of the generated photon or SPP field which are registered by the single-photon detectors.


{\it Generation of entangled surface plasmon-polaritons}. The proposed formalism can be easily extended to the generation of SPPs owing to the universality of the dyadic Green's functions. For a nanoparticle located in the vicinity of an interface, dyadic Green's function is given by the sum of two contributions: $\hat{G} = \hat{G}^{(0)} + \hat{G}^{({\rm ref})}$. The vacuum term $\hat{G}^{(0)}$ corresponds to the dipole radiation in free space and is given by the well-known expression \cite{Novotny_2012} $G^{(0)}_{\alpha \beta}(\mathbf{r},\mathbf{r}', \omega) = \frac{{\rm exp}(ik_0R)}{4\pi R}\Big[\Big( \delta_{\alpha \beta} - \frac{R_{\alpha}R_{\beta}}{R^2}\Big) - \frac{1 - ik_0R}{k_{0}^{2}R^2}\Big(\delta_{\alpha \beta} - 3\frac{R_{\alpha}R_{\beta}}{R^{2}}\Big)\Big]$, where $\mathbf{R}=\mathbf{r}-\mathbf{r}'$, $R = |\mathbf{R}|$, and $k_0 = \omega/c$ is a wave vector in the free space. The second contribution $\hat{G}^{({\rm ref})}$ takes into account reflection from the interface, and can be represented in the form of a single integral \cite{2019_Kostina}
\begin{equation}
   \hat{G}^{({\rm ref})}(\rho, \varphi, z) = \frac{i k_0}{8\pi}\int_0^{\infty}\hat{M}(s, \rho, \varphi){\rm exp}(i s_{1z} z)ds.
   \label{eq:G_Integral}
\end{equation}
Here the variable $s(\omega) = \sqrt{k_{x}^{2}(\omega) + k_{y}^{2}(\omega)}/k_{0}(\omega)$ is the dimensionless wave vector, and $s_{z}^{\rm Ag,air}(\omega) =\sqrt{\varepsilon_{\rm Ag,air}(\omega) - s(\omega)}$ is its $z$-component in the substrate or air, respectively. With the considered $z$-polarized detectors, the following elements of the  matrix $\hat{M}$ are required:
\begin{equation*}
\begin{split}
   M_{xz} &= -M_{zx} = -2\pi i s^{2}r_{p}J_{1}(s\rho){\rm cos}(\varphi),\\
   M_{yz} &= -M_{zy} = -2\pi i s^{2}r_{p}J_{1}(s\rho){\rm sin}(\varphi),\\
   M_{zz} &= 2\pi J_{0}(s\rho)r_{p}\frac{s^{3}}{s_{1z}},
\end{split}
\end{equation*}
where $J_{0,1}(s\rho)$ are Bessel functions of the first kind and $r_p(\omega) = (\varepsilon_{\rm Ag}(\omega) s_{z}^{\rm air} - \varepsilon_{\rm air}s_{z}^{\rm Ag}/(\varepsilon_{\rm Ag}(\omega)s_{z}^{\rm air} + \varepsilon_{\rm air} s_{z}^{\rm Ag})$ is the Fresnel reflection coefficient, with $\varepsilon_{\rm air}$ and $\varepsilon_{\rm Ag}(\omega)$ being permittivities of the air and silver substrate, respectively. The permittivity of the substrate $\varepsilon_{\rm Ag}(\omega)$ includes both real and imaginary parts, thus taking into account ohmic losses and related dissipation, as well as details of the surface plasmon-polariton dispersion. In the case of purely imaginary $s_{z}^{\rm air}$, characteristic of evanescent waves localized at the surface, the term $\hat{G}^{({\rm ref})}$ describes the electric field of SPP's. To evaluate the corresponding integrals Eq.(\ref{eq:G_Integral}) numerically, we use experimentally measured values of the silver permittivity from Ref.~\cite{1972_Johnson}.

\section{Results and Discussion}
\label{sec:Results}

{\it Generation of SPP ${\rm N00N}$ states with a $\text{GaAs}$ nanoparticle}. Hereafter, we consider a system composed of a ${\rm GaAs}$ nanoparticle located at the silver-air interface, Fig.~\ref{fig:Geometry}. Non-zero elements of the corresponding $\chi^{(2)}$ tensor are $\chi^{(2)}_{xyz} = \chi^{(2)}_{zxy} = \chi^{(2)}_{yzx} = \chi^{(2)}_{yxz} = \chi^{(2)}_{zyx} = \chi^{(2)}_{xzy} = \chi^{(2)}_{0}$~\cite{Boyd_2008}. We assume that the size of the nanoparticle $a$ is related to the pump radiation wavelength $\lambda$ and the refractive index of the nanoparticle material $n$ as $a \ll \lambda/n$, and the nanoparticle can be treated as a point electric dipole. Thus, we consider a case when the particle itself is non-resonant, and all the resonant behavior is attributed to the excitation of surface plasmon-polaritons. From now on, we focus our attention on the generation of SPP pairs.

\begin{figure}[b]
    \centering
    \includegraphics[width=8.5cm]{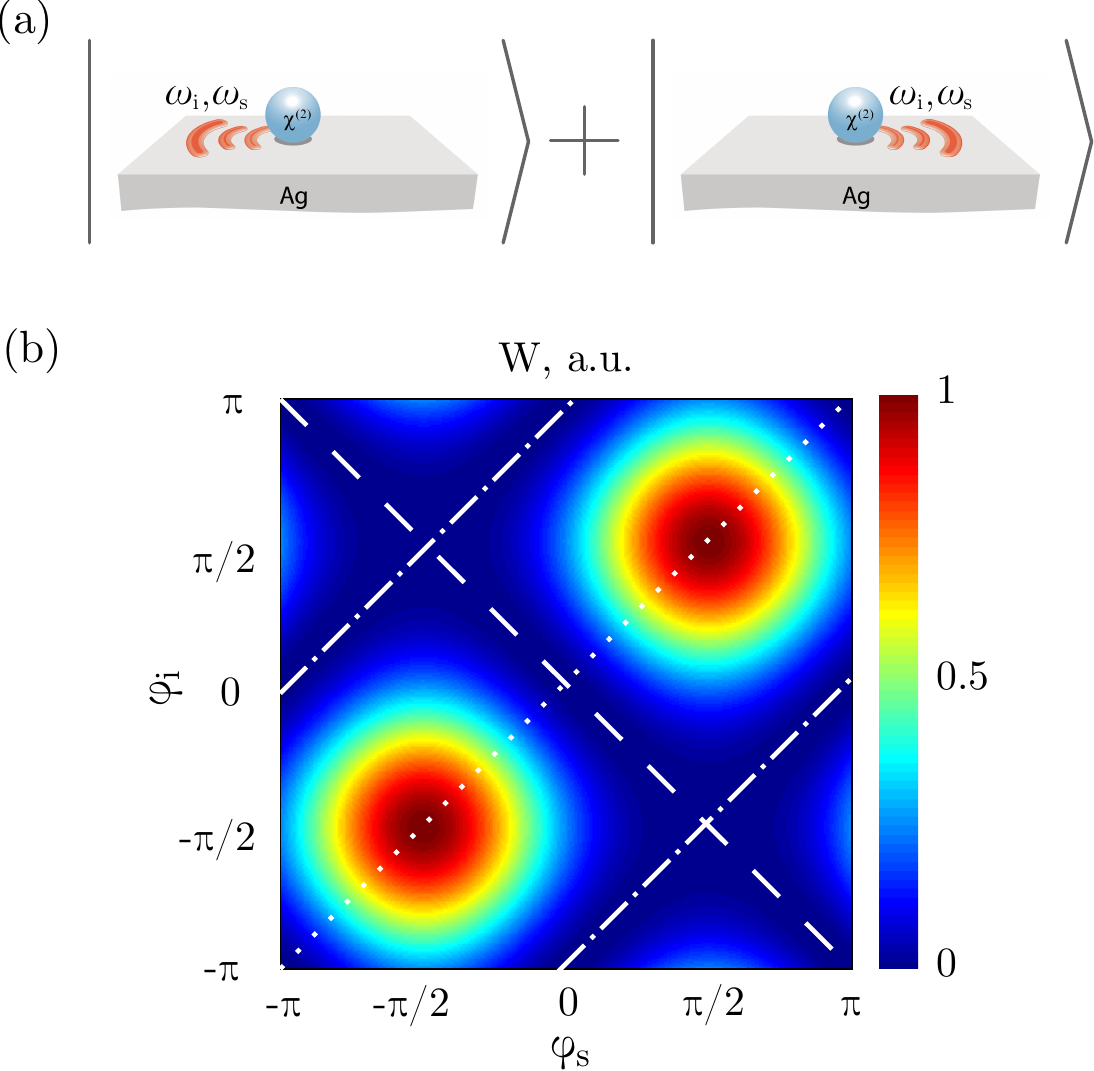}
    \caption{(a) Schematic representation of the generation of a ${\rm N00N}$-state of surface plasmon-polariton pair during the SPDC process. (b) Two-plasmon counting rate $W$ as a function of in-plane angular positions of detectors $\varphi_{\rm i}$, $\varphi_{\rm s}$ spanning the entire $2\pi$ range. The angles $\varphi_{\rm i}$ and $\varphi_{\rm s}$ are calculated with respect to the positive direction of the pump wave's electric field. The following parameters are used: $\theta=0^{\rm o}$, $a=30~{\rm nm}$, $z_{\rm i} = z_{\rm s} = 15~{\rm nm}$, $\rho_{\rm i} = \rho_{\rm s} = 12~{\rm \mu m}$.}
    \label{fig:N00N}
\end{figure}

We consider the following geometry of the system. A nanoparticle made of ${\rm GaAs}$ is placed at the silver-air interface in such a way that crystalline axes of the material are directed along the $x||[001]$, $y||[010]$ and $z||[100]$ axes of the Cartesian system shown in  Fig.~\ref{fig:Geometry}. The pump is linearly polarized in the ${\rm TM}$-geometry corresponding to the electric field of the pump wave located in the $xz$-plane. At normal incidence, $\theta=0^{\rm o}$, the pump electric field is directed along the $x$-axis. Then, the only two non-zero elements of the generation matrix $\Gamma_{\alpha\beta} = E_0\chi^{(2)}_{\alpha\beta x}e_{x}$ are $\Gamma_{yz}$ and $\Gamma_{zy}$, and the corresponding dyadic decomposition reads $\hat{\Gamma} = E_0\chi^{(2)}(\ket{e_{y}}\bra{e_{z}} + \ket{e_{z}}\bra{e_{y}})$. For convenience, we consider both detectors located at the same radial distance $\rho_{0}$ in the $xy$-plane with respect to the center of the nanoparticle projection onto the mentioned plane, and at the same height $z_{0}$. In our numerical simulations, we take the values $\rho_{0}=12~{\rm \mu m}$ and $z_{0}=10~{\rm nm}$ which correspond to the far-field region as well as to the detection of surface waves with electric field concentrated in the vicinity of the interface. Then, the coordinate dependence of the amplitude Eq.~(\ref{eq:T}) reduces to the angular positions of the detectors $\varphi_{\rm i}$ and $\varphi_{\rm s}$ only, $T(\mathbf{r}_{\rm i}, \mathbf{r}_{\rm s}) = T(\varphi_{i}, \varphi_{s})$, as shown in Fig.~\ref{fig:Geometry}.

The ${\rm N00N}$-state, Figure~\ref{fig:N00N}(a), is one of the four Bell states having the following form in the case of $N=2$:
\begin{equation}
    \ket{\psi} = \frac{1}{\sqrt{2}}\Big(\ket{u}_{i}\ket{u}_{s} + \ket{v}_{i}\ket{v}_{s}\Big),
    \label{eq:N00N_Definition}
\end{equation}
To characterize the degree of entanglement for the generated states, we calculate Schmidt number \cite{1995_Ekert, 2004_Law}
\begin{equation}
    K = \frac{\sum\limits_m \Lambda_m^2}{\sum\limits_m \Lambda_m^4}
    \label{eq:Schmidt}
\end{equation}
where $\Lambda_m$ are Schmidt coefficients defined via a decomposition of two-plasmon wavefunction $T(\varphi_{\rm i}, \varphi_{\rm s})$ in the basis of single-plasmon modes $u^{(\rm{i})}(\varphi_{\rm i})$ and $v^{(\rm{s})}(\varphi_{\rm s})$ as $\textstyle T = \sum\limits_{m=1}^N \sqrt{\Lambda_m}u^{(\rm{i})}_{m}\times v^{(\rm{s})}_{m}$. As can be demonstrated by performing a required singular value decomposition of the two-photon wave function $T(\varphi_{\rm i}, \varphi_{\rm s})$, single-photon basis functions represent angular parts of dipole Green's functions depicted in Fig.~\ref{fig:Induced_Dipoles}: ${\rm Re}\{u^{(\rm{i})}_{1}({\varphi_{\rm i}})\}, {\rm Im}\{u^{(\rm{i})}_{1}({\varphi_{\rm i}})\} \propto {\rm cos}({\varphi_{\rm i}})$, ${\rm Re}\{v^{(\rm{s})}_{1}({\varphi_{\rm s}})\}, {\rm Im}\{v^{(\rm{s})}_{1}({\varphi_{\rm s}})\} \propto {\rm sin}({\varphi_{\rm s}})$, and vice versa for $m=2$. In contrast to detection rate $W$, Schmidt number takes into account phase of the two-photon wave function $T$, thus allowing more rigorous classification of quantum states. For maximally entangled states, Schmidt number is $K=2$, whereas for non-entangled states $K=1$ \cite{2004_Law}. However, the criteria $K=2$ is not sufficient to guarantee the maximally entangled state in a system of identical bosons \cite{2004_Ghirardi}, thus the consideration of the corresponding figure of merit is needed.

\begin{figure}[t]
    \centering
    \includegraphics[width=8.5cm]{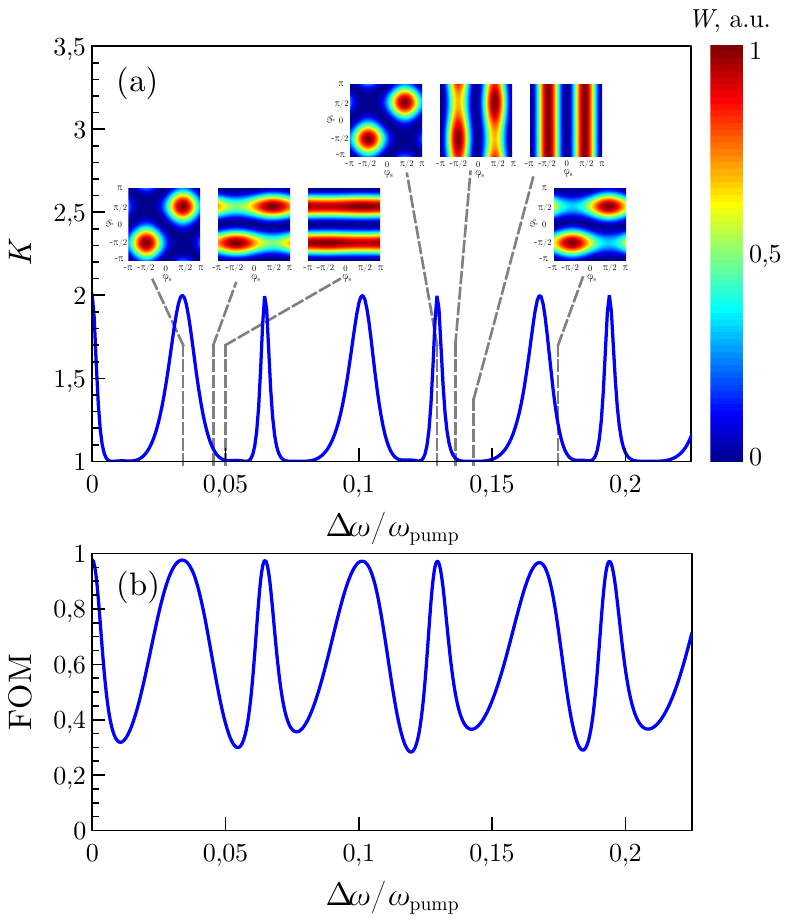}
    \caption{(a): Schmidt number $K$ Eq.~(\ref{eq:Schmidt}) as a function of frequency difference $\Delta\omega = \omega_{\rm s} - \omega_{\rm i}$ for $\lambda_{\rm pump} = 750 {\rm nm}$ and $\theta_{\rm pump} = 0^{\rm o}$. Insets  show corresponding detection rate patterns at frequency differences marked with dashed lines. (b): Figure of merit Eq.(\ref{eq:FOM}) as a function of frequency difference $\Delta\omega$ for the same set of parameters as in (a).}
    \label{fig:Schmidt}
\end{figure}

To do so, we define the partial two-plasmon wavefunctions $T_{\rm L}$ and $T_{\rm R}$ as follows: $T_{\rm L}(\varphi_{\rm i},\varphi_{\rm s}) = T(\varphi_{\rm i},\varphi_{\rm s})$ for $0\le \phi_{\rm i,s} \le \pi$, and  $T_{\rm L}=0$ for $-\pi\le \phi_{\rm i,s} \le 0$; $T_{\rm R}(\varphi_{\rm i},\varphi_{\rm s}) = T(\varphi_{\rm i},\varphi_{\rm s})$ for  $-\pi\le \phi_{\rm i,s} \le 0$, and  $T_{\rm L}=0$ for $0\le \phi_{\rm i,s} \le \pi$. Then, we perform separate Schmidt decompositions for the amplitudes $T_{\rm L}$ and $T_{\rm R}$, obtaining Schmidt coefficients $\Lambda_{\rm L1}$, $\Lambda_{\rm R1}$ for $m=1$, and the corresponding single-plasmon modes $u_{\rm L1}^{(\rm i)}$, $u_{\rm R1}^{(\rm i)}$ for the idler plasmon, and $v_{\rm L1}^{(\rm s)}$, $v_{\rm R1}^{(\rm s)}$ for the signal plasmon, respectively. Next, we introduce the set of partial figures of merit (FOM) with $P_{\rm T} = \int_{-\pi}^{\pi}\int_{-\pi}^{\pi}|T(\varphi_{\rm i}, \omega_{\rm i}, \varphi_{\rm s}, \omega_{\rm s})|^{2}{\rm d}\varphi_{\rm i}{\rm d}\varphi_{\rm s}$ introduced for normalization. The single-plasmon wavefunctions should be the same for idler and signal plasmons in the ${\rm N00N}$-state Eq.(\ref{eq:N00N_Definition}), causing ${\rm  FOM}_{m_{\rm L}} = \Big|\bra{u_{\rm L1}^{(\rm i)}}\ket{ v_{\rm L1}^{(\rm s)}}\Big|^{2} = {\rm  FOM}_{m_{\rm R}} = \Big|\bra{u_{\rm R1}^{(\rm i)}}\ket{ v_{\rm R1}^{(\rm s)}}\Big|^{2} = 1$. The next two conditions ${\rm  FOM}_{\rm i} = 1 - \Big|\bra{u_{\rm L1}^{(\rm i)}}\ket{u_{\rm R1}^{(\rm i)}}\Big|^{2} = 1$ and ${\rm  FOM}_{\rm s} = 1 - \Big|\bra{v_{\rm L1}^{(\rm s)}}\ket{v_{\rm R1}^{(\rm s)}}\Big|^{2} = 1$ require the orthogonality of single-plasmon wavefunctions for the amplitudes $T_{\rm L}$ and $T_{\rm R}$, also required for the state Eq.(\ref{eq:N00N_Definition}). Finally, for the ${\rm N00N}$-state $\Lambda_{\rm L1}^{2}/P_{\rm T} = \Lambda_{\rm R1}^{2}/P_{\rm T} = 1$, and ${\rm  FOM}_{\Lambda_{\rm L}} =  1 - \Big(\Lambda_{\rm L1}^2\frac{2}{P_{\rm T}} - 1\Big)^{2} = {\rm  FOM}_{\Lambda_{\rm R}} = 1 - \Big(\Lambda_{\rm R1}^2\frac{2}{P_{\rm T}} - 1\Big)^{2} = 1$. Thus, we define the resulting figure of merit as follows:
\begin{equation}
\begin{split}
    {\rm  FOM} = {\rm  FOM}_{\Lambda_{\rm L}}&\cdot{\rm  FOM}_{\Lambda_{\rm R}}\cdot\\
    &\cdot{\rm  FOM}_{m_{\rm L}}{\rm  FOM}_{m_{\rm R}}{\rm  FOM}_{i}{\rm  FOM}_{s}.
    \label{eq:FOM}
\end{split}
\end{equation}
Such FOM equals $1$ for the ${\rm N00N}$-state, and ${\rm FOM}<1$ corresponds to states not satisfying the definition Eq.(\ref{eq:N00N_Definition}).

By setting the polarizations of idler and signal detectors to be linear and directed along the $z$-axis, $\mathbf{d}_{\rm i} = \mathbf{d}_{\rm s} = d_{0}\mathbf{e}_z$, we obtain a simple expression for the two-plasmon amplitude:
\begin{equation}
    T(\varphi_{i}, \varphi_{s})/T_0 = G_{zy}^{(i)}G_{zz}^{(s)} + G_{zz}^{(i)}G_{yz}^{(s)},
    \label{eq:T_N00N}
\end{equation}
where the dimensional prefactor is $T_0 = \chi^{(2)}_{0}E^{\rm pump}d_{0}^{2}$ and the indices $(i)$, $(s)$ denote Green's functions with the arguments $(\varphi_{\rm i}, \omega_{\rm i})$ and $(\mathbf{r}_{\rm s}, \mathbf{r}_{\rm 0}, \omega_{\rm s})$, respectively. The combination of the Green's function components with account of their symmetry $G_{\alpha\beta}^{(i, s)}=G_{\beta\alpha}^{(i,s)}$ ensures the pattern demonstrated in Fig.~\ref{fig:N00N}(b). This can be demonstrated by tracking the magnitude of the two-plasmon detection rate $W(\varphi_{\rm i}, \varphi_{\rm s})\propto |T|^{2}$ along several characteristic lines marked in Fig.~\ref{fig:N00N}(b). Indeed, for $\varphi_{\rm i}=\varphi_{\rm s}$ the Green's functions for the signal and idler plasmons coincide, $G_{\alpha\beta}^{({\rm i})} = G_{\alpha\beta}^{({\rm s})} = G_{\alpha\beta}$, and thus the expression for the two-photon counting rate takes the form $W/W_0 = 4(G_{zy}^{'2}G_{zz}^{'2} + G_{zy}^{''2}G_{zz}^{''2} + G_{zy}^{''2}G_{zz}^{'2} + G_{zy}^{'2}G_{zz}^{''2}) = 4(G_{zz}^{'2} + G_{zz}^{''2})(G_{zy}^{'2} + G_{zy}^{''2})$, where single and double primes denote the real and imaginary parts of the Green's functions, correspondingly, and the prefactor is $W_{0}=\frac{2\pi}{\hbar}|T_0|^2$. As seen from Fig.~\ref{fig:N00N}(b), this quantity reaches its maximum at $\varphi_{\rm i,s}=\pm \pi/2$, whereas for $\varphi=0$ and $\varphi=\pm\pi$ two-plasmon detection rate equals zero. This profile is shown with the dotted line in Fig.~\ref{fig:N00N}(b). At the same time, for $\varphi_{\rm i} = \varphi_{\rm s} + \pi$ the idler and signal Green's functions relate as $G_{zz}^{({\rm i})}=G_{zz}^{({\rm s})}$, $G_{zy}^{({\rm i})}=-G_{zy}^{({\rm s})}$ which results in $W(\varphi, \varphi+\pi)=0$ (dashed line in Fig.~\ref{fig:N00N}(b)). The same holds for $\varphi_{\rm i} = -\varphi_{\rm s}$, hence $W(\varphi, -\varphi)=0$ as well (dash-dotted line in Fig.~\ref{fig:N00N}(b)). The direct evaluation of Eq.(\ref{eq:Schmidt}) and  Eq.(\ref{eq:FOM}) for the obtained state gives $K=2$ and ${\rm FOM}=1$. Thus, the state in Fig.~\ref{fig:N00N}(b) is a ${\rm N00N}$-state generated as a result of a nonlinear interference related to both the properties of the $\chi^{(2)}$ tensor of the nanoparticle as well as to the general properties of Green's function for the dipole source exciting surface waves.


\begin{figure}[t]
    \centering
    \includegraphics[width=8.5cm]{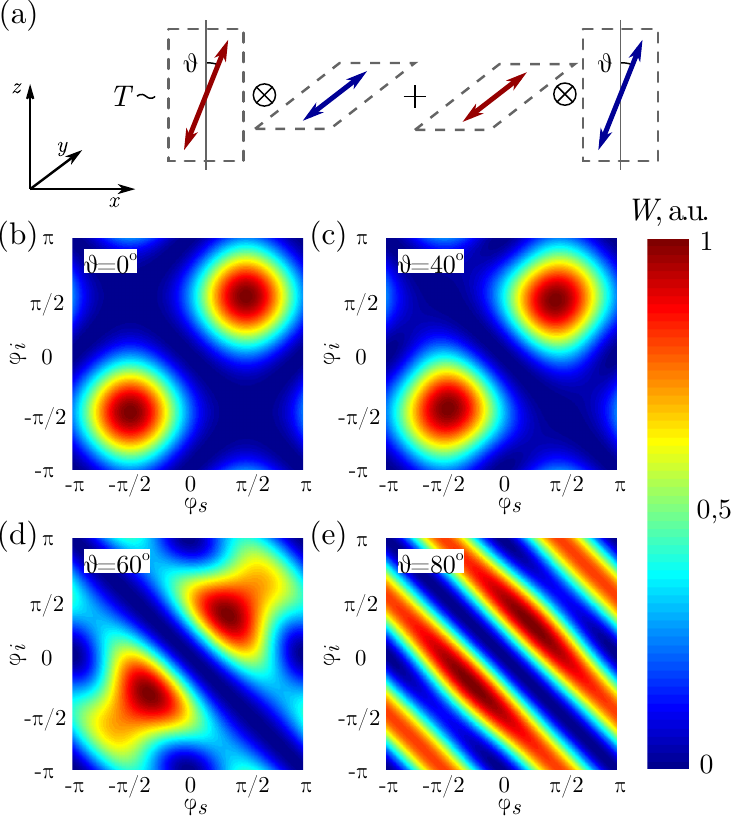}
    \caption{(a) Induced dipole decomposition of the two-photon wave function Eq.\ref{eq:T} in the general case of TM-polarized pump at the incidence angle $\theta$. (b-e) Two-plasmon detection rates $W$ for the same parameters of the system as in Fig.\ref{fig:Geometry}(c), but at different pump angles $\theta$ within the TM-geometry.}
    \label{fig:N00N_Angular}
\end{figure}

{\it Robustness of the ${\rm N00N}$-state generation}. Next, we consider a non-degenerate SPDC process with different frequencies of the detected plasmons $\omega_{\rm i} \neq \omega_{\rm s}$ (see Fig.~\ref{fig:Schmidt}(a)). In such a case, Schmidt number $K$ plotted as a function of the frequency difference between the idler and signal SPPs $\Delta\omega$ oscillates between $K=1$ corresponding to factorizable states, and $K=2$, Fig.~\ref{fig:Schmidt}(a). Similar oscillations are observed in the FOM, Fig.~\ref{fig:Schmidt}(b). Together these quantities highlight the spectral positions of the entangled ${\rm N00N}$-states. Such oscillating behavior results from the difference in the wavelengths of idler and signal plasmons, which affects the interference conditions of the idler and signal Green's functions in Eq.~(\ref{eq:T_N00N}). Thus, a considerable entanglement can be obtained either for a slightly non-degenerate SPDC with the difference between idler and signal frequencies $\Delta\omega < 1\%$, or for a strongly non-degenerate SPDC with frequencies $\omega_{i}$ and $\omega_{s}$ differing by respective discrete steps seen in Fig.~\ref{fig:Schmidt}(a,b). These results should also be taken into account when performing a post-selection of the generated SPP pairs.

Finally, we examine the robustness of the ${\rm N00N}$-state generation with respect to changes in the pump incidence angle $\theta$. It can be straightforwardly obtained, that in this case one of the induced dipoles is oriented at the angle $\theta$ with respect to the normal in the $xz$-plane, whereas another one is still oriented along the $y$-axis, Fig.~\ref{fig:N00N_Angular}(a). In Figs.~\ref{fig:N00N_Angular}(b-e) the two-plasmon detection rates at different angles of the pump photon incidence $\theta$ within the TM-geometry are shown. It is seen that at the angle $\theta=40^{\rm o}$, the pattern remains nearly unchanged and still features pronounced maxima for both plasmons generated either at the in-plane angles $\varphi_{i,s}=-\pi/2$ or $\varphi_{i,s}=\pi/2$, Fig.~\ref{fig:N00N_Angular}(c). However, if the pump angle $\theta$ differs from $0^{\rm o}$, then the dyadic decomposition Eq.~(\ref{eq:Gamma}) includes additional terms having $x$-projection that change the desired ${\rm N00N}$-pattern. In particular, for $\theta=60^{\rm o}$, two characteristic maxima are readily smoothen with these additional contributions, Fig.~\ref{fig:N00N_Angular}(d). Finally, at oblique incidence $\theta=80^{\rm o}$ the two-plasmon amplitude represents a set of stripes $\varphi_{s}=\varphi_{i} \pm \pi/2$ corresponding to the equally probable generation of one plasmon in two opposite directions, both being orthogonal to the direction of the second plasmon, Fig.~\ref{fig:N00N_Angular}(e). Thus, the ${\rm N00N}$-state generation remains robust even at pump angles changed by $30^{\rm o}..40^{\rm o}$ from normal.

\section{Conclusion}
\label{sec:Conclusion}

In this article, we studied the generation of surface plasmon-polariton pairs through spontaneous parametric down-conversion taking place in a $\chi^{(2)}$-nanoparticle located in the vicinity of a metallic substrate. We have revealed that a ${\rm N00N}$-state generation can be achieved with a ${\rm GaAs}$ nanoparticle which is robust towards changes in the pump incidence angle and pump frequency, and can be observed between certain frequencies of idler and signal plasmons.

The obtained ${\rm N00N}$-pattern is related mostly to the general form of dyadic Green's function of dipole emitter exciting surface waves, and to the form of the $\chi^{(2)}$ tensor in combination with the excitation and detection geometry. Thus, the obtained results and developed approach can be straightforwardly generalized to more sophisticated cases of nanoparticle assemblies that preserve the desired symmetries of Green's function. Other possible extensions include the generation of various surface waves besides surface plasmon-polaritons, and generating ${\rm N00N}$-states of photon pairs by outcoupling the emitted plasmons in the far-field. This renders the considered scheme as a general recipe for the nonlinear generation of ${\rm N00N}$-states with nanoparticles.

\section*{Acknowledgments}
We are grateful to Alexander Poddubny and Daria Smirnova for fruitful discussions. The work is financially supported by the Russian Foundation for Basic Research (proj. 18-32-01052, 18-02-01206 and 18-29-20037) and the Australian Research Council (DE180100070, DP160100619, DP190101559, CE200100010). N.O. acknowledges partial support by the Foundation for the Advancement of Theoretical Physics and Mathematics ``BASIS''.


\begin{thebibliography}{99}
\makeatletter
\providecommand \@ifxundefined [1]{%
 \@ifx{#1\undefined}
}%
\providecommand \@ifnum [1]{%
 \ifnum #1\expandafter \@firstoftwo
 \else \expandafter \@secondoftwo
 \fi
}%
\providecommand \@ifx [1]{%
 \ifx #1\expandafter \@firstoftwo
 \else \expandafter \@secondoftwo
 \fi
}%
\providecommand \natexlab [1]{#1}%
\providecommand \enquote  [1]{``#1''}%
\providecommand \bibnamefont  [1]{#1}%
\providecommand \bibfnamefont [1]{#1}%
\providecommand \citenamefont [1]{#1}%
\providecommand \href@noop [0]{\@secondoftwo}%
\providecommand \href [0]{\begingroup \@sanitize@url \@href}%
\providecommand \@href[1]{\@@startlink{#1}\@@href}%
\providecommand \@@href[1]{\endgroup#1\@@endlink}%
\providecommand \@sanitize@url [0]{\catcode `\\12\catcode `\$12\catcode
  `\&12\catcode `\#12\catcode `\^12\catcode `\_12\catcode `\%12\relax}%
\providecommand \@@startlink[1]{}%
\providecommand \@@endlink[0]{}%
\providecommand \url  [0]{\begingroup\@sanitize@url \@url }%
\providecommand \@url [1]{\endgroup\@href {#1}{\urlprefix }}%
\providecommand \urlprefix  [0]{URL }%
\providecommand \Eprint [0]{\href }%
\providecommand \doibase [0]{http://dx.doi.org/}%
\providecommand \selectlanguage [0]{\@gobble}%
\providecommand \bibinfo  [0]{\@secondoftwo}%
\providecommand \bibfield  [0]{\@secondoftwo}%
\providecommand \translation [1]{[#1]}%
\providecommand \BibitemOpen [0]{}%
\providecommand \bibitemStop [0]{}%
\providecommand \bibitemNoStop [0]{.\EOS\space}%
\providecommand \EOS [0]{\spacefactor3000\relax}%
\providecommand \BibitemShut  [1]{\csname bibitem#1\endcsname}%
\let\auto@bib@innerbib\@empty
\bibitem [{\citenamefont {M\"{u}ller}\ \emph {et~al.}(2014)\citenamefont
  {M\"{u}ller}, \citenamefont {Bounouar}, \citenamefont {J\"{o}ns},
  \citenamefont {Gl\"{a}ssl},\ and\ \citenamefont {Michler}}]{2014_Muller}%
  \BibitemOpen
  \bibfield  {author} {\bibinfo {author} {\bibfnamefont {M.}~\bibnamefont
  {M\"{u}ller}}, \bibinfo {author} {\bibfnamefont {S.}~\bibnamefont
  {Bounouar}}, \bibinfo {author} {\bibfnamefont {K.~D.}\ \bibnamefont
  {J\"{o}ns}}, \bibinfo {author} {\bibfnamefont {M.}~\bibnamefont
  {Gl\"{a}ssl}}, \ and\ \bibinfo {author} {\bibfnamefont {P.}~\bibnamefont
  {Michler}},\ }\bibfield  {title} {\enquote {\bibinfo {title} {On-demand
  generation of indistinguishable polarization-entangled photon pairs},}\
  }\href {\doibase 10.1038/nphoton.2013.377} {\bibfield  {journal} {\bibinfo
  {journal} {Nature Photonics}\ }\textbf {\bibinfo {volume} {8}},\ \bibinfo
  {pages} {224--228} (\bibinfo {year} {2014})}\BibitemShut {NoStop}%
\bibitem [{\citenamefont {Versteegh}\ \emph {et~al.}(2014)\citenamefont
  {Versteegh}, \citenamefont {Reimer}, \citenamefont {J\"{o}ns}, \citenamefont
  {Dalacu}, \citenamefont {Poole}, \citenamefont {Gulinatti}, \citenamefont
  {Giudice},\ and\ \citenamefont {Zwiller}}]{2014_Versteegh}%
  \BibitemOpen
  \bibfield  {author} {\bibinfo {author} {\bibfnamefont {Marijn A.~M.}\
  \bibnamefont {Versteegh}}, \bibinfo {author} {\bibfnamefont {Michael~E.}\
  \bibnamefont {Reimer}}, \bibinfo {author} {\bibfnamefont {Klaus~D.}\
  \bibnamefont {J\"{o}ns}}, \bibinfo {author} {\bibfnamefont {Dan}\
  \bibnamefont {Dalacu}}, \bibinfo {author} {\bibfnamefont {Philip~J.}\
  \bibnamefont {Poole}}, \bibinfo {author} {\bibfnamefont {Angelo}\
  \bibnamefont {Gulinatti}}, \bibinfo {author} {\bibfnamefont {Andrea}\
  \bibnamefont {Giudice}}, \ and\ \bibinfo {author} {\bibfnamefont {Val}\
  \bibnamefont {Zwiller}},\ }\bibfield  {title} {\enquote {\bibinfo {title}
  {Observation of strongly entangled photon pairs from a nanowire quantum
  dot},}\ }\href {\doibase 10.1038/ncomms6298} {\bibfield  {journal} {\bibinfo
  {journal} {Nature Communications}\ }\textbf {\bibinfo {volume} {5}},\
  \bibinfo {pages} {5298} (\bibinfo {year} {2014})}\BibitemShut {NoStop}%
\bibitem [{\citenamefont {Solntsev}\ and\ \citenamefont
  {Sukhorukov}(2017)}]{2017_Solntsev}%
  \BibitemOpen
  \bibfield  {author} {\bibinfo {author} {\bibfnamefont {Alexander~S.}\
  \bibnamefont {Solntsev}}\ and\ \bibinfo {author} {\bibfnamefont {Andrey~A.}\
  \bibnamefont {Sukhorukov}},\ }\bibfield  {title} {\enquote {\bibinfo {title}
  {Path-entangled photon sources on nonlinear chips},}\ }\href {\doibase
  10.1016/j.revip.2016.11.003} {\bibfield  {journal} {\bibinfo  {journal}
  {Reviews in Physics}\ }\textbf {\bibinfo {volume} {2}},\ \bibinfo {pages}
  {19--31} (\bibinfo {year} {2017})}\BibitemShut {NoStop}%
\bibitem [{\citenamefont {O'Brien}\ \emph {et~al.}(2009)\citenamefont
  {O'Brien}, \citenamefont {Furusawa},\ and\ \citenamefont
  {Vu{\v{c}}kovi{\'{c}}}}]{2009_OBrien}%
  \BibitemOpen
  \bibfield  {author} {\bibinfo {author} {\bibfnamefont {Jeremy~L.}\
  \bibnamefont {O'Brien}}, \bibinfo {author} {\bibfnamefont {Akira}\
  \bibnamefont {Furusawa}}, \ and\ \bibinfo {author} {\bibfnamefont {Jelena}\
  \bibnamefont {Vu{\v{c}}kovi{\'{c}}}},\ }\bibfield  {title} {\enquote
  {\bibinfo {title} {Photonic quantum technologies},}\ }\href {\doibase
  10.1038/nphoton.2009.229} {\bibfield  {journal} {\bibinfo  {journal} {Nature
  Photonics}\ }\textbf {\bibinfo {volume} {3}},\ \bibinfo {pages} {687--695}
  (\bibinfo {year} {2009})}\BibitemShut {NoStop}%
\bibitem [{\citenamefont {Solntsev}\ \emph {et~al.}(2018)\citenamefont
  {Solntsev}, \citenamefont {Kumar}, \citenamefont {Pertsch}, \citenamefont
  {Sukhorukov},\ and\ \citenamefont {Setzpfandt}}]{2018_Solntsev}%
  \BibitemOpen
  \bibfield  {author} {\bibinfo {author} {\bibfnamefont {Alexander~S.}\
  \bibnamefont {Solntsev}}, \bibinfo {author} {\bibfnamefont {Pawan}\
  \bibnamefont {Kumar}}, \bibinfo {author} {\bibfnamefont {Thomas}\
  \bibnamefont {Pertsch}}, \bibinfo {author} {\bibfnamefont {Andrey~A.}\
  \bibnamefont {Sukhorukov}}, \ and\ \bibinfo {author} {\bibfnamefont {Frank}\
  \bibnamefont {Setzpfandt}},\ }\bibfield  {title} {\enquote {\bibinfo {title}
  {{LiNbO}3waveguides for integrated {SPDC} spectroscopy},}\ }\href {\doibase
  10.1063/1.5009766} {\bibfield  {journal} {\bibinfo  {journal} {{APL}
  Photonics}\ }\textbf {\bibinfo {volume} {3}},\ \bibinfo {pages} {021301}
  (\bibinfo {year} {2018})}\BibitemShut {NoStop}%
\bibitem [{\citenamefont {Fakonas}\ \emph {et~al.}(2014)\citenamefont
  {Fakonas}, \citenamefont {Lee}, \citenamefont {Kelaita},\ and\ \citenamefont
  {Atwater}}]{2014_Fakonas}%
  \BibitemOpen
  \bibfield  {author} {\bibinfo {author} {\bibfnamefont {James~S.}\
  \bibnamefont {Fakonas}}, \bibinfo {author} {\bibfnamefont {Hyunseok}\
  \bibnamefont {Lee}}, \bibinfo {author} {\bibfnamefont {Yousif~A.}\
  \bibnamefont {Kelaita}}, \ and\ \bibinfo {author} {\bibfnamefont {Harry~A.}\
  \bibnamefont {Atwater}},\ }\bibfield  {title} {\enquote {\bibinfo {title}
  {Two-plasmon quantum interference},}\ }\href {\doibase
  10.1038/nphoton.2014.40} {\bibfield  {journal} {\bibinfo  {journal} {Nature
  Photonics}\ }\textbf {\bibinfo {volume} {8}},\ \bibinfo {pages} {317--320}
  (\bibinfo {year} {2014})}\BibitemShut {NoStop}%
\bibitem [{\citenamefont {Dheur}\ \emph {et~al.}(2016)\citenamefont {Dheur},
  \citenamefont {Devaux}, \citenamefont {Ebbesen}, \citenamefont {Baron},
  \citenamefont {Rodier}, \citenamefont {Hugonin}, \citenamefont {Lalanne},
  \citenamefont {Greffet}, \citenamefont {Messin},\ and\ \citenamefont
  {Marquier}}]{2016_Dheur}%
  \BibitemOpen
  \bibfield  {author} {\bibinfo {author} {\bibfnamefont {Marie-Christine}\
  \bibnamefont {Dheur}}, \bibinfo {author} {\bibfnamefont {Eloïse}\
  \bibnamefont {Devaux}}, \bibinfo {author} {\bibfnamefont {Thomas~W.}\
  \bibnamefont {Ebbesen}}, \bibinfo {author} {\bibfnamefont {Alexandre}\
  \bibnamefont {Baron}}, \bibinfo {author} {\bibfnamefont {Jean-Claude}\
  \bibnamefont {Rodier}}, \bibinfo {author} {\bibfnamefont {Jean-Paul}\
  \bibnamefont {Hugonin}}, \bibinfo {author} {\bibfnamefont {Philippe}\
  \bibnamefont {Lalanne}}, \bibinfo {author} {\bibfnamefont {Jean-Jacques}\
  \bibnamefont {Greffet}}, \bibinfo {author} {\bibfnamefont {Ga{\'{e}}tan}\
  \bibnamefont {Messin}}, \ and\ \bibinfo {author} {\bibfnamefont
  {Fran{\c{c}}ois}\ \bibnamefont {Marquier}},\ }\bibfield  {title} {\enquote
  {\bibinfo {title} {Single-plasmon interferences},}\ }\href {\doibase
  10.1126/sciadv.1501574} {\bibfield  {journal} {\bibinfo  {journal} {Science
  Advances}\ }\textbf {\bibinfo {volume} {2}},\ \bibinfo {pages} {e1501574}
  (\bibinfo {year} {2016})}\BibitemShut {NoStop}%
\bibitem [{\citenamefont {Dheur}\ \emph {et~al.}(2017)\citenamefont {Dheur},
  \citenamefont {Vest}, \citenamefont {Devaux}, \citenamefont {Baron},
  \citenamefont {Hugonin}, \citenamefont {Greffet}, \citenamefont {Messin},\
  and\ \citenamefont {Marquier}}]{2017_Dheur}%
  \BibitemOpen
  \bibfield  {author} {\bibinfo {author} {\bibfnamefont {Marie-Christine}\
  \bibnamefont {Dheur}}, \bibinfo {author} {\bibfnamefont {Benjamin}\
  \bibnamefont {Vest}}, \bibinfo {author} {\bibfnamefont {{\'{E}}loïse}\
  \bibnamefont {Devaux}}, \bibinfo {author} {\bibfnamefont {Alexandre}\
  \bibnamefont {Baron}}, \bibinfo {author} {\bibfnamefont {Jean-Paul}\
  \bibnamefont {Hugonin}}, \bibinfo {author} {\bibfnamefont {Jean-Jacques}\
  \bibnamefont {Greffet}}, \bibinfo {author} {\bibfnamefont {Ga{\'{e}}tan}\
  \bibnamefont {Messin}}, \ and\ \bibinfo {author} {\bibfnamefont
  {Fran{\c{c}}ois}\ \bibnamefont {Marquier}},\ }\bibfield  {title} {\enquote
  {\bibinfo {title} {Remote preparation of single-plasmon states},}\ }\href
  {\doibase 10.1103/physrevb.96.045432} {\bibfield  {journal} {\bibinfo
  {journal} {Physical Review B}\ }\textbf {\bibinfo {volume} {96}},\ \bibinfo
  {pages} {045432} (\bibinfo {year} {2017})}\BibitemShut {NoStop}%
\bibitem [{\citenamefont {D'Amico}\ \emph {et~al.}(2019)\citenamefont
  {D'Amico}, \citenamefont {Angelakis}, \citenamefont {Bussi{\`{e}}res},
  \citenamefont {Caglayan}, \citenamefont {Couteau}, \citenamefont {Durt},
  \citenamefont {Kolaric}, \citenamefont {Maletinsky}, \citenamefont
  {Pfeiffer}, \citenamefont {Rabl}, \citenamefont {Xuereb},\ and\ \citenamefont
  {Agio}}]{2019_DAmico}%
  \BibitemOpen
  \bibfield  {author} {\bibinfo {author} {\bibfnamefont {Irene}\ \bibnamefont
  {D'Amico}}, \bibinfo {author} {\bibfnamefont {Dimitris~G.}\ \bibnamefont
  {Angelakis}}, \bibinfo {author} {\bibfnamefont {F{\'{e}}lix}\ \bibnamefont
  {Bussi{\`{e}}res}}, \bibinfo {author} {\bibfnamefont {Humeyra}\ \bibnamefont
  {Caglayan}}, \bibinfo {author} {\bibfnamefont {Christophe}\ \bibnamefont
  {Couteau}}, \bibinfo {author} {\bibfnamefont {Thomas}\ \bibnamefont {Durt}},
  \bibinfo {author} {\bibfnamefont {Branko}\ \bibnamefont {Kolaric}}, \bibinfo
  {author} {\bibfnamefont {Patrick}\ \bibnamefont {Maletinsky}}, \bibinfo
  {author} {\bibfnamefont {Walter}\ \bibnamefont {Pfeiffer}}, \bibinfo {author}
  {\bibfnamefont {Peter}\ \bibnamefont {Rabl}}, \bibinfo {author}
  {\bibfnamefont {Andr{\'{e}}}\ \bibnamefont {Xuereb}}, \ and\ \bibinfo
  {author} {\bibfnamefont {Mario}\ \bibnamefont {Agio}},\ }\bibfield  {title}
  {\enquote {\bibinfo {title} {Nanoscale quantum optics},}\ }\href {\doibase
  10.1393/ncr/i2019-10158-0} {\bibfield  {journal} {\bibinfo  {journal} {La
  Rivista del Nuovo Cimento}\ }\textbf {\bibinfo {volume} {42}},\ \bibinfo
  {pages} {153--195} (\bibinfo {year} {2019})}\BibitemShut {NoStop}%
\bibitem [{\citenamefont {Dowling}\ and\ \citenamefont
  {Seshadreesan}(2015)}]{2015_Dowling}%
  \BibitemOpen
  \bibfield  {author} {\bibinfo {author} {\bibfnamefont {Jonathan~P.}\
  \bibnamefont {Dowling}}\ and\ \bibinfo {author} {\bibfnamefont {Kaushik~P.}\
  \bibnamefont {Seshadreesan}},\ }\bibfield  {title} {\enquote {\bibinfo
  {title} {Quantum optical technologies for metrology, sensing, and imaging},}\
  }\href {http://jlt.osa.org/abstract.cfm?URI=jlt-33-12-2359} {\bibfield
  {journal} {\bibinfo  {journal} {J. Lightwave Technol.}\ }\textbf {\bibinfo
  {volume} {33}},\ \bibinfo {pages} {2359--2370} (\bibinfo {year}
  {2015})}\BibitemShut {NoStop}%
\bibitem [{\citenamefont {Kannan}\ \emph {et~al.}(2020)\citenamefont {Kannan},
  \citenamefont {Campbell}, \citenamefont {Vasconcelos}, \citenamefont {Winik},
  \citenamefont {Kim}, \citenamefont {Kjaergaard}, \citenamefont {Krantz},
  \citenamefont {Melville}, \citenamefont {Niedzielski}, \citenamefont {Yoder},
  \citenamefont {Orlando}, \citenamefont {Gustavsson},\ and\ \citenamefont
  {Oliver}}]{2020_Kannan}%
  \BibitemOpen
  \bibfield  {author} {\bibinfo {author} {\bibfnamefont {B.}~\bibnamefont
  {Kannan}}, \bibinfo {author} {\bibfnamefont {D.~L.}\ \bibnamefont
  {Campbell}}, \bibinfo {author} {\bibfnamefont {F.}~\bibnamefont
  {Vasconcelos}}, \bibinfo {author} {\bibfnamefont {R.}~\bibnamefont {Winik}},
  \bibinfo {author} {\bibfnamefont {D.~K.}\ \bibnamefont {Kim}}, \bibinfo
  {author} {\bibfnamefont {M.}~\bibnamefont {Kjaergaard}}, \bibinfo {author}
  {\bibfnamefont {P.}~\bibnamefont {Krantz}}, \bibinfo {author} {\bibfnamefont
  {A.}~\bibnamefont {Melville}}, \bibinfo {author} {\bibfnamefont {B.~M.}\
  \bibnamefont {Niedzielski}}, \bibinfo {author} {\bibfnamefont {J.~L.}\
  \bibnamefont {Yoder}}, \bibinfo {author} {\bibfnamefont {T.~P.}\ \bibnamefont
  {Orlando}}, \bibinfo {author} {\bibfnamefont {S.}~\bibnamefont {Gustavsson}},
  \ and\ \bibinfo {author} {\bibfnamefont {W.~D.}\ \bibnamefont {Oliver}},\
  }\bibfield  {title} {\enquote {\bibinfo {title} {Generating spatially
  entangled itinerant photons with waveguide quantum electrodynamics},}\ }\href
  {\doibase 10.1126/sciadv.abb8780} {\bibfield  {journal} {\bibinfo  {journal}
  {Science Advances}\ }\textbf {\bibinfo {volume} {6}},\ \bibinfo {pages}
  {eabb8780} (\bibinfo {year} {2020})}\BibitemShut {NoStop}%
\bibitem [{\citenamefont {Rechtsman}\ \emph {et~al.}(2016)\citenamefont
  {Rechtsman}, \citenamefont {Lumer}, \citenamefont {Plotnik}, \citenamefont
  {Perez-Leija}, \citenamefont {Szameit},\ and\ \citenamefont
  {Segev}}]{2016_Rechtsman}%
  \BibitemOpen
  \bibfield  {author} {\bibinfo {author} {\bibfnamefont {Mikael~C.}\
  \bibnamefont {Rechtsman}}, \bibinfo {author} {\bibfnamefont {Yaakov}\
  \bibnamefont {Lumer}}, \bibinfo {author} {\bibfnamefont {Yonatan}\
  \bibnamefont {Plotnik}}, \bibinfo {author} {\bibfnamefont {Armando}\
  \bibnamefont {Perez-Leija}}, \bibinfo {author} {\bibfnamefont {Alexander}\
  \bibnamefont {Szameit}}, \ and\ \bibinfo {author} {\bibfnamefont {Mordechai}\
  \bibnamefont {Segev}},\ }\bibfield  {title} {\enquote {\bibinfo {title}
  {Topological protection of photonic path entanglement},}\ }\href {\doibase
  10.1364/optica.3.000925} {\bibfield  {journal} {\bibinfo  {journal} {Optica}\
  }\textbf {\bibinfo {volume} {3}},\ \bibinfo {pages} {925} (\bibinfo {year}
  {2016})}\BibitemShut {NoStop}%
\bibitem [{\citenamefont {Han}\ \emph {et~al.}(2020)\citenamefont {Han},
  \citenamefont {Sukhorukov},\ and\ \citenamefont {Leykam}}]{2020_Han}%
  \BibitemOpen
  \bibfield  {author} {\bibinfo {author} {\bibfnamefont {JungYun}\ \bibnamefont
  {Han}}, \bibinfo {author} {\bibfnamefont {Andrey~A.}\ \bibnamefont
  {Sukhorukov}}, \ and\ \bibinfo {author} {\bibfnamefont {Daniel}\ \bibnamefont
  {Leykam}},\ }\bibfield  {title} {\enquote {\bibinfo {title}
  {Disorder-protected quantum state transmission through helical
  coupled-resonator waveguides},}\ }\href {\doibase 10.1364/PRJ.399919}
  {\bibfield  {journal} {\bibinfo  {journal} {Photon. Res.}\ }\textbf {\bibinfo
  {volume} {8}},\ \bibinfo {pages} {B15--B24} (\bibinfo {year}
  {2020})}\BibitemShut {NoStop}%
\bibitem [{\citenamefont {Chen}\ \emph {et~al.}(2018)\citenamefont {Chen},
  \citenamefont {Lee}, \citenamefont {Lu}, \citenamefont {Liu}, \citenamefont
  {Wu}, \citenamefont {Feng}, \citenamefont {Li}, \citenamefont {Rockstuhl},
  \citenamefont {Guo}, \citenamefont {Guo}, \citenamefont {Tame},\ and\
  \citenamefont {Ren}}]{2018_Chen}%
  \BibitemOpen
  \bibfield  {author} {\bibinfo {author} {\bibfnamefont {Yang}\ \bibnamefont
  {Chen}}, \bibinfo {author} {\bibfnamefont {Changhyoup}\ \bibnamefont {Lee}},
  \bibinfo {author} {\bibfnamefont {Liu}\ \bibnamefont {Lu}}, \bibinfo {author}
  {\bibfnamefont {Di}~\bibnamefont {Liu}}, \bibinfo {author} {\bibfnamefont
  {Yun-Kun}\ \bibnamefont {Wu}}, \bibinfo {author} {\bibfnamefont {Lan-Tian}\
  \bibnamefont {Feng}}, \bibinfo {author} {\bibfnamefont {Ming}\ \bibnamefont
  {Li}}, \bibinfo {author} {\bibfnamefont {Carsten}\ \bibnamefont {Rockstuhl}},
  \bibinfo {author} {\bibfnamefont {Guo-Ping}\ \bibnamefont {Guo}}, \bibinfo
  {author} {\bibfnamefont {Guang-Can}\ \bibnamefont {Guo}}, \bibinfo {author}
  {\bibfnamefont {Mark}\ \bibnamefont {Tame}}, \ and\ \bibinfo {author}
  {\bibfnamefont {Xi-Feng}\ \bibnamefont {Ren}},\ }\bibfield  {title} {\enquote
  {\bibinfo {title} {Quantum plasmonic {N00N} state in a silver nanowire and
  its use for quantum sensing},}\ }\href {\doibase 10.1364/optica.5.001229}
  {\bibfield  {journal} {\bibinfo  {journal} {Optica}\ }\textbf {\bibinfo
  {volume} {5}},\ \bibinfo {pages} {1229} (\bibinfo {year} {2018})}\BibitemShut
  {NoStop}%
\bibitem [{\citenamefont {Vest}\ \emph {et~al.}(2018)\citenamefont {Vest},
  \citenamefont {Shlesinger}, \citenamefont {Dheur}, \citenamefont {Devaux},
  \citenamefont {Greffet}, \citenamefont {Messin},\ and\ \citenamefont
  {Marquier}}]{2018_Vest}%
  \BibitemOpen
  \bibfield  {author} {\bibinfo {author} {\bibfnamefont {Benjamin}\
  \bibnamefont {Vest}}, \bibinfo {author} {\bibfnamefont {Ilan}\ \bibnamefont
  {Shlesinger}}, \bibinfo {author} {\bibfnamefont {Marie-Christine}\
  \bibnamefont {Dheur}}, \bibinfo {author} {\bibfnamefont {{\'{E}}loïse}\
  \bibnamefont {Devaux}}, \bibinfo {author} {\bibfnamefont {Jean-Jacques}\
  \bibnamefont {Greffet}}, \bibinfo {author} {\bibfnamefont {Ga{\'{e}}tan}\
  \bibnamefont {Messin}}, \ and\ \bibinfo {author} {\bibfnamefont
  {Fran{\c{c}}ois}\ \bibnamefont {Marquier}},\ }\bibfield  {title} {\enquote
  {\bibinfo {title} {Plasmonic interferences of two-particle {N00N} states},}\
  }\href {\doibase 10.1088/1367-2630/aac24f} {\bibfield  {journal} {\bibinfo
  {journal} {New Journal of Physics}\ }\textbf {\bibinfo {volume} {20}},\
  \bibinfo {pages} {053050} (\bibinfo {year} {2018})}\BibitemShut {NoStop}%
\bibitem [{\citenamefont {Mehta}\ \emph {et~al.}(2020)\citenamefont {Mehta},
  \citenamefont {Achanta},\ and\ \citenamefont {Dasgupta}}]{2020_Mehta}%
  \BibitemOpen
  \bibfield  {author} {\bibinfo {author} {\bibfnamefont {Karun}\ \bibnamefont
  {Mehta}}, \bibinfo {author} {\bibfnamefont {Venu~Gopal}\ \bibnamefont
  {Achanta}}, \ and\ \bibinfo {author} {\bibfnamefont {Shubhrangshu}\
  \bibnamefont {Dasgupta}},\ }\bibfield  {title} {\enquote {\bibinfo {title}
  {Generation of non-classical states of photons from a
  metal{\textendash}dielectric interface: a novel architecture for quantum
  information processing},}\ }\href {\doibase 10.1039/c9nr06529f} {\bibfield
  {journal} {\bibinfo  {journal} {Nanoscale}\ }\textbf {\bibinfo {volume}
  {12}},\ \bibinfo {pages} {256--261} (\bibinfo {year} {2020})}\BibitemShut
  {NoStop}%
\bibitem [{\citenamefont {Lee}\ \emph {et~al.}(2018)\citenamefont {Lee},
  \citenamefont {Yoon}, \citenamefont {Rah}, \citenamefont {Tame},
  \citenamefont {Rockstuhl}, \citenamefont {Song}, \citenamefont {Lee},\ and\
  \citenamefont {Lee}}]{2018_Lee}%
  \BibitemOpen
  \bibfield  {author} {\bibinfo {author} {\bibfnamefont {Joong-Sung}\
  \bibnamefont {Lee}}, \bibinfo {author} {\bibfnamefont {Seung-Jin}\
  \bibnamefont {Yoon}}, \bibinfo {author} {\bibfnamefont {Hyungju}\
  \bibnamefont {Rah}}, \bibinfo {author} {\bibfnamefont {Mark}\ \bibnamefont
  {Tame}}, \bibinfo {author} {\bibfnamefont {Carsten}\ \bibnamefont
  {Rockstuhl}}, \bibinfo {author} {\bibfnamefont {Seok~Ho}\ \bibnamefont
  {Song}}, \bibinfo {author} {\bibfnamefont {Changhyoup}\ \bibnamefont {Lee}},
  \ and\ \bibinfo {author} {\bibfnamefont {Kwang-Geol}\ \bibnamefont {Lee}},\
  }\bibfield  {title} {\enquote {\bibinfo {title} {Quantum plasmonic sensing
  using single photons},}\ }\href {\doibase 10.1364/oe.26.029272} {\bibfield
  {journal} {\bibinfo  {journal} {Optics Express}\ }\textbf {\bibinfo {volume}
  {26}},\ \bibinfo {pages} {29272} (\bibinfo {year} {2018})}\BibitemShut
  {NoStop}%
\bibitem [{\citenamefont {Fan}\ \emph {et~al.}(2015)\citenamefont {Fan},
  \citenamefont {Lawrie},\ and\ \citenamefont {Pooser}}]{2015_Fan}%
  \BibitemOpen
  \bibfield  {author} {\bibinfo {author} {\bibfnamefont {Wenjiang}\
  \bibnamefont {Fan}}, \bibinfo {author} {\bibfnamefont {Benjamin~J.}\
  \bibnamefont {Lawrie}}, \ and\ \bibinfo {author} {\bibfnamefont {Raphael~C.}\
  \bibnamefont {Pooser}},\ }\bibfield  {title} {\enquote {\bibinfo {title}
  {Quantum plasmonic sensing},}\ }\href {\doibase 10.1103/physreva.92.053812}
  {\bibfield  {journal} {\bibinfo  {journal} {Physical Review A}\ }\textbf
  {\bibinfo {volume} {92}},\ \bibinfo {pages} {053812} (\bibinfo {year}
  {2015})}\BibitemShut {NoStop}%
\bibitem [{\citenamefont {Lawrie}\ \emph {et~al.}(2019)\citenamefont {Lawrie},
  \citenamefont {Lett}, \citenamefont {Marino},\ and\ \citenamefont
  {Pooser}}]{2019_Lawrie}%
  \BibitemOpen
  \bibfield  {author} {\bibinfo {author} {\bibfnamefont {B.~J.}\ \bibnamefont
  {Lawrie}}, \bibinfo {author} {\bibfnamefont {P.~D.}\ \bibnamefont {Lett}},
  \bibinfo {author} {\bibfnamefont {A.~M.}\ \bibnamefont {Marino}}, \ and\
  \bibinfo {author} {\bibfnamefont {R.~C.}\ \bibnamefont {Pooser}},\ }\bibfield
   {title} {\enquote {\bibinfo {title} {Quantum sensing with squeezed light},}\
  }\href {\doibase 10.1021/acsphotonics.9b00250} {\bibfield  {journal}
  {\bibinfo  {journal} {{ACS} Photonics}\ }\textbf {\bibinfo {volume} {6}},\
  \bibinfo {pages} {1307--1318} (\bibinfo {year} {2019})}\BibitemShut {NoStop}%
\bibitem [{\citenamefont {Camacho-Morales}\ \emph {et~al.}(2016)\citenamefont
  {Camacho-Morales}, \citenamefont {Rahmani}, \citenamefont {Kruk},
  \citenamefont {Wang}, \citenamefont {Xu}, \citenamefont {Smirnova},
  \citenamefont {Solntsev}, \citenamefont {Miroshnichenko}, \citenamefont
  {Tan}, \citenamefont {Karouta}, \citenamefont {Naureen}, \citenamefont
  {Vora}, \citenamefont {Carletti}, \citenamefont {Angelis}, \citenamefont
  {Jagadish}, \citenamefont {Kivshar},\ and\ \citenamefont
  {Neshev}}]{2016_Camacho_Morales}%
  \BibitemOpen
  \bibfield  {author} {\bibinfo {author} {\bibfnamefont {Rocio}\ \bibnamefont
  {Camacho-Morales}}, \bibinfo {author} {\bibfnamefont {Mohsen}\ \bibnamefont
  {Rahmani}}, \bibinfo {author} {\bibfnamefont {Sergey}\ \bibnamefont {Kruk}},
  \bibinfo {author} {\bibfnamefont {Lei}\ \bibnamefont {Wang}}, \bibinfo
  {author} {\bibfnamefont {Lei}\ \bibnamefont {Xu}}, \bibinfo {author}
  {\bibfnamefont {Daria~A.}\ \bibnamefont {Smirnova}}, \bibinfo {author}
  {\bibfnamefont {Alexander~S.}\ \bibnamefont {Solntsev}}, \bibinfo {author}
  {\bibfnamefont {Andrey}\ \bibnamefont {Miroshnichenko}}, \bibinfo {author}
  {\bibfnamefont {Hark~Hoe}\ \bibnamefont {Tan}}, \bibinfo {author}
  {\bibfnamefont {Fouad}\ \bibnamefont {Karouta}}, \bibinfo {author}
  {\bibfnamefont {Shagufta}\ \bibnamefont {Naureen}}, \bibinfo {author}
  {\bibfnamefont {Kaushal}\ \bibnamefont {Vora}}, \bibinfo {author}
  {\bibfnamefont {Luca}\ \bibnamefont {Carletti}}, \bibinfo {author}
  {\bibfnamefont {Costantino~De}\ \bibnamefont {Angelis}}, \bibinfo {author}
  {\bibfnamefont {Chennupati}\ \bibnamefont {Jagadish}}, \bibinfo {author}
  {\bibfnamefont {Yuri~S.}\ \bibnamefont {Kivshar}}, \ and\ \bibinfo {author}
  {\bibfnamefont {Dragomir~N.}\ \bibnamefont {Neshev}},\ }\bibfield  {title}
  {\enquote {\bibinfo {title} {Nonlinear generation of vector beams from
  {AlGaAs} nanoantennas},}\ }\href {\doibase 10.1021/acs.nanolett.6b03525}
  {\bibfield  {journal} {\bibinfo  {journal} {Nano Letters}\ }\textbf {\bibinfo
  {volume} {16}},\ \bibinfo {pages} {7191--7197} (\bibinfo {year}
  {2016})}\BibitemShut {NoStop}%
\bibitem [{\citenamefont {Carletti}\ \emph {et~al.}(2017)\citenamefont
  {Carletti}, \citenamefont {Rocco}, \citenamefont {Locatelli}, \citenamefont
  {Angelis}, \citenamefont {Gili}, \citenamefont {Ravaro}, \citenamefont
  {Favero}, \citenamefont {Leo}, \citenamefont {Finazzi}, \citenamefont
  {Ghirardini}, \citenamefont {Celebrano}, \citenamefont {Marino},\ and\
  \citenamefont {Zayats}}]{2017_Carletti}%
  \BibitemOpen
  \bibfield  {author} {\bibinfo {author} {\bibfnamefont {L}~\bibnamefont
  {Carletti}}, \bibinfo {author} {\bibfnamefont {D}~\bibnamefont {Rocco}},
  \bibinfo {author} {\bibfnamefont {A}~\bibnamefont {Locatelli}}, \bibinfo
  {author} {\bibfnamefont {C~De}\ \bibnamefont {Angelis}}, \bibinfo {author}
  {\bibfnamefont {V~F}\ \bibnamefont {Gili}}, \bibinfo {author} {\bibfnamefont
  {M}~\bibnamefont {Ravaro}}, \bibinfo {author} {\bibfnamefont {I}~\bibnamefont
  {Favero}}, \bibinfo {author} {\bibfnamefont {G}~\bibnamefont {Leo}}, \bibinfo
  {author} {\bibfnamefont {M}~\bibnamefont {Finazzi}}, \bibinfo {author}
  {\bibfnamefont {L}~\bibnamefont {Ghirardini}}, \bibinfo {author}
  {\bibfnamefont {M}~\bibnamefont {Celebrano}}, \bibinfo {author}
  {\bibfnamefont {G}~\bibnamefont {Marino}}, \ and\ \bibinfo {author}
  {\bibfnamefont {A~V}\ \bibnamefont {Zayats}},\ }\bibfield  {title} {\enquote
  {\bibinfo {title} {Controlling second-harmonic generation at the nanoscale
  with monolithic {AlGaAs}-on-{AlOx} antennas},}\ }\href {\doibase
  10.1088/1361-6528/aa5645} {\bibfield  {journal} {\bibinfo  {journal}
  {Nanotechnology}\ }\textbf {\bibinfo {volume} {28}},\ \bibinfo {pages}
  {114005} (\bibinfo {year} {2017})}\BibitemShut {NoStop}%
\bibitem [{\citenamefont {Liu}\ \emph {et~al.}(2018)\citenamefont {Liu},
  \citenamefont {Vabishchevich}, \citenamefont {Vaskin}, \citenamefont {Reno},
  \citenamefont {Keeler}, \citenamefont {Sinclair}, \citenamefont {Staude},\
  and\ \citenamefont {Brener}}]{2018_Liu}%
  \BibitemOpen
  \bibfield  {author} {\bibinfo {author} {\bibfnamefont {Sheng}\ \bibnamefont
  {Liu}}, \bibinfo {author} {\bibfnamefont {Polina~P.}\ \bibnamefont
  {Vabishchevich}}, \bibinfo {author} {\bibfnamefont {Aleksandr}\ \bibnamefont
  {Vaskin}}, \bibinfo {author} {\bibfnamefont {John~L.}\ \bibnamefont {Reno}},
  \bibinfo {author} {\bibfnamefont {Gordon~A.}\ \bibnamefont {Keeler}},
  \bibinfo {author} {\bibfnamefont {Michael~B.}\ \bibnamefont {Sinclair}},
  \bibinfo {author} {\bibfnamefont {Isabelle}\ \bibnamefont {Staude}}, \ and\
  \bibinfo {author} {\bibfnamefont {Igal}\ \bibnamefont {Brener}},\ }\bibfield
  {title} {\enquote {\bibinfo {title} {An all-dielectric metasurface as a
  broadband optical frequency mixer},}\ }\href {\doibase
  10.1038/s41467-018-04944-9} {\bibfield  {journal} {\bibinfo  {journal}
  {Nature Communications}\ }\textbf {\bibinfo {volume} {9}},\ \bibinfo {pages}
  {2507} (\bibinfo {year} {2018})}\BibitemShut {NoStop}%
\bibitem [{\citenamefont {Saerens}\ \emph {et~al.}(2020)\citenamefont
  {Saerens}, \citenamefont {Tang}, \citenamefont {Petrov}, \citenamefont
  {Frizyuk}, \citenamefont {Renaut}, \citenamefont {Timpu}, \citenamefont
  {Escal{\'{e}}}, \citenamefont {Shtrom}, \citenamefont {Bouravleuv},
  \citenamefont {Cirlin}, \citenamefont {Grange},\ and\ \citenamefont
  {Timofeeva}}]{2020_Saerens}%
  \BibitemOpen
  \bibfield  {author} {\bibinfo {author} {\bibfnamefont {Gr{\'{e}}goire}\
  \bibnamefont {Saerens}}, \bibinfo {author} {\bibfnamefont {Iek}\ \bibnamefont
  {Tang}}, \bibinfo {author} {\bibfnamefont {Mihail~I.}\ \bibnamefont
  {Petrov}}, \bibinfo {author} {\bibfnamefont {Kristina}\ \bibnamefont
  {Frizyuk}}, \bibinfo {author} {\bibfnamefont {Claude}\ \bibnamefont
  {Renaut}}, \bibinfo {author} {\bibfnamefont {Flavia}\ \bibnamefont {Timpu}},
  \bibinfo {author} {\bibfnamefont {Marc~Reig}\ \bibnamefont {Escal{\'{e}}}},
  \bibinfo {author} {\bibfnamefont {Igor}\ \bibnamefont {Shtrom}}, \bibinfo
  {author} {\bibfnamefont {Alexey}\ \bibnamefont {Bouravleuv}}, \bibinfo
  {author} {\bibfnamefont {George}\ \bibnamefont {Cirlin}}, \bibinfo {author}
  {\bibfnamefont {Rachel}\ \bibnamefont {Grange}}, \ and\ \bibinfo {author}
  {\bibfnamefont {Maria}\ \bibnamefont {Timofeeva}},\ }\bibfield  {title}
  {\enquote {\bibinfo {title} {Engineering of the second-harmonic emission
  directionality with {III}{\textendash}v semiconductor rod nanoantennas},}\
  }\href {\doibase 10.1002/lpor.202000028} {\bibfield  {journal} {\bibinfo
  {journal} {Laser {\&} Photonics Reviews}\ }\textbf {\bibinfo {volume} {14}},\
  \bibinfo {pages} {2000028} (\bibinfo {year} {2020})}\BibitemShut {NoStop}%
\bibitem [{\citenamefont {Marino}\ \emph {et~al.}(2019)\citenamefont {Marino},
  \citenamefont {Solntsev}, \citenamefont {Xu}, \citenamefont {Gili},
  \citenamefont {Carletti}, \citenamefont {Poddubny}, \citenamefont {Rahmani},
  \citenamefont {Smirnova}, \citenamefont {Chen}, \citenamefont
  {Lema{\^{\i}}tre}, \citenamefont {Zhang}, \citenamefont {Zayats},
  \citenamefont {Angelis}, \citenamefont {Leo}, \citenamefont {Sukhorukov},\
  and\ \citenamefont {Neshev}}]{2019_Marino}%
  \BibitemOpen
  \bibfield  {author} {\bibinfo {author} {\bibfnamefont {Giuseppe}\
  \bibnamefont {Marino}}, \bibinfo {author} {\bibfnamefont {Alexander~S.}\
  \bibnamefont {Solntsev}}, \bibinfo {author} {\bibfnamefont {Lei}\
  \bibnamefont {Xu}}, \bibinfo {author} {\bibfnamefont {Valerio~F.}\
  \bibnamefont {Gili}}, \bibinfo {author} {\bibfnamefont {Luca}\ \bibnamefont
  {Carletti}}, \bibinfo {author} {\bibfnamefont {Alexander~N.}\ \bibnamefont
  {Poddubny}}, \bibinfo {author} {\bibfnamefont {Mohsen}\ \bibnamefont
  {Rahmani}}, \bibinfo {author} {\bibfnamefont {Daria~A.}\ \bibnamefont
  {Smirnova}}, \bibinfo {author} {\bibfnamefont {Haitao}\ \bibnamefont {Chen}},
  \bibinfo {author} {\bibfnamefont {Aristide}\ \bibnamefont {Lema{\^{\i}}tre}},
  \bibinfo {author} {\bibfnamefont {Guoquan}\ \bibnamefont {Zhang}}, \bibinfo
  {author} {\bibfnamefont {Anatoly~V.}\ \bibnamefont {Zayats}}, \bibinfo
  {author} {\bibfnamefont {Costantino~De}\ \bibnamefont {Angelis}}, \bibinfo
  {author} {\bibfnamefont {Giuseppe}\ \bibnamefont {Leo}}, \bibinfo {author}
  {\bibfnamefont {Andrey~A.}\ \bibnamefont {Sukhorukov}}, \ and\ \bibinfo
  {author} {\bibfnamefont {Dragomir~N.}\ \bibnamefont {Neshev}},\ }\bibfield
  {title} {\enquote {\bibinfo {title} {Spontaneous photon-pair generation from
  a dielectric nanoantenna},}\ }\href {\doibase 10.1364/optica.6.001416}
  {\bibfield  {journal} {\bibinfo  {journal} {Optica}\ }\textbf {\bibinfo
  {volume} {6}},\ \bibinfo {pages} {1416} (\bibinfo {year} {2019})}\BibitemShut
  {NoStop}%
\bibitem [{\citenamefont {Okoth}\ \emph {et~al.}(2019)\citenamefont {Okoth},
  \citenamefont {Cavanna}, \citenamefont {Santiago-Cruz},\ and\ \citenamefont
  {Chekhova}}]{2019_Okoth}%
  \BibitemOpen
  \bibfield  {author} {\bibinfo {author} {\bibfnamefont {C.}~\bibnamefont
  {Okoth}}, \bibinfo {author} {\bibfnamefont {A.}~\bibnamefont {Cavanna}},
  \bibinfo {author} {\bibfnamefont {T.}~\bibnamefont {Santiago-Cruz}}, \ and\
  \bibinfo {author} {\bibfnamefont {M.{\hspace{0.167em}}V.}\ \bibnamefont
  {Chekhova}},\ }\bibfield  {title} {\enquote {\bibinfo {title} {Microscale
  generation of entangled photons without momentum conservation},}\ }\href
  {\doibase 10.1103/physrevlett.123.263602} {\bibfield  {journal} {\bibinfo
  {journal} {Physical Review Letters}\ }\textbf {\bibinfo {volume} {123}},\
  \bibinfo {pages} {263602} (\bibinfo {year} {2019})}\BibitemShut {NoStop}%
\bibitem [{\citenamefont {Santiago-Cruz}\ \emph {et~al.}(2020)\citenamefont
  {Santiago-Cruz}, \citenamefont {Sultanov}, \citenamefont {Zhang},
  \citenamefont {Krivitsky},\ and\ \citenamefont
  {Chekhova}}]{2020_Santiago_Cruz}%
  \BibitemOpen
  \bibfield  {author} {\bibinfo {author} {\bibfnamefont {Tomás}\ \bibnamefont
  {Santiago-Cruz}}, \bibinfo {author} {\bibfnamefont {Vitaliy}\ \bibnamefont
  {Sultanov}}, \bibinfo {author} {\bibfnamefont {Haizhong}\ \bibnamefont
  {Zhang}}, \bibinfo {author} {\bibfnamefont {Leonid~A.}\ \bibnamefont
  {Krivitsky}}, \ and\ \bibinfo {author} {\bibfnamefont {Maria~V.}\
  \bibnamefont {Chekhova}},\ }\href@noop {} {\enquote {\bibinfo {title}
  {Spontaneous parametric down-conversion from subwavelength nonlinear
  films},}\ } (\bibinfo {year} {2020}),\ \Eprint
  {http://arxiv.org/abs/2009.00324} {arXiv:2009.00324 [quant-ph]} \BibitemShut
  {NoStop}%
\bibitem [{\citenamefont {Li}\ \emph {et~al.}(2020)\citenamefont {Li},
  \citenamefont {Liu}, \citenamefont {Ren}, \citenamefont {Wang}, \citenamefont
  {Su}, \citenamefont {Chen}, \citenamefont {Chu}, \citenamefont {Kuo},
  \citenamefont {Liu}, \citenamefont {Zang}, \citenamefont {Guo}, \citenamefont
  {Zhang}, \citenamefont {Wang}, \citenamefont {Zhu},\ and\ \citenamefont
  {Tsai}}]{2020_Li}%
  \BibitemOpen
  \bibfield  {author} {\bibinfo {author} {\bibfnamefont {Lin}\ \bibnamefont
  {Li}}, \bibinfo {author} {\bibfnamefont {Zexuan}\ \bibnamefont {Liu}},
  \bibinfo {author} {\bibfnamefont {Xifeng}\ \bibnamefont {Ren}}, \bibinfo
  {author} {\bibfnamefont {Shuming}\ \bibnamefont {Wang}}, \bibinfo {author}
  {\bibfnamefont {Vin-Cent}\ \bibnamefont {Su}}, \bibinfo {author}
  {\bibfnamefont {Mu-Ku}\ \bibnamefont {Chen}}, \bibinfo {author}
  {\bibfnamefont {Cheng~Hung}\ \bibnamefont {Chu}}, \bibinfo {author}
  {\bibfnamefont {Hsin~Yu}\ \bibnamefont {Kuo}}, \bibinfo {author}
  {\bibfnamefont {Biheng}\ \bibnamefont {Liu}}, \bibinfo {author}
  {\bibfnamefont {Wenbo}\ \bibnamefont {Zang}}, \bibinfo {author}
  {\bibfnamefont {Guangcan}\ \bibnamefont {Guo}}, \bibinfo {author}
  {\bibfnamefont {Lijian}\ \bibnamefont {Zhang}}, \bibinfo {author}
  {\bibfnamefont {Zhenlin}\ \bibnamefont {Wang}}, \bibinfo {author}
  {\bibfnamefont {Shining}\ \bibnamefont {Zhu}}, \ and\ \bibinfo {author}
  {\bibfnamefont {Din~Ping}\ \bibnamefont {Tsai}},\ }\bibfield  {title}
  {\enquote {\bibinfo {title} {Metalens-array{\textendash}based
  high-dimensional and multiphoton quantum source},}\ }\href {\doibase
  10.1126/science.aba9779} {\bibfield  {journal} {\bibinfo  {journal}
  {Science}\ }\textbf {\bibinfo {volume} {368}},\ \bibinfo {pages} {1487--1490}
  (\bibinfo {year} {2020})}\BibitemShut {NoStop}%
\bibitem [{\citenamefont {Santiago-Cruz}\ \emph {et~al.}(2021)\citenamefont
  {Santiago-Cruz}, \citenamefont {Fedotova}, \citenamefont {Sultanov},
  \citenamefont {Weissflog}, \citenamefont {Arslan}, \citenamefont {Younesi},
  \citenamefont {Pertsch}, \citenamefont {Staude}, \citenamefont {Setzpfandt},\
  and\ \citenamefont {Chekhova}}]{2021_Santiago_Cruz}%
  \BibitemOpen
  \bibfield  {author} {\bibinfo {author} {\bibfnamefont {Tomás}\ \bibnamefont
  {Santiago-Cruz}}, \bibinfo {author} {\bibfnamefont {Anna}\ \bibnamefont
  {Fedotova}}, \bibinfo {author} {\bibfnamefont {Vitaliy}\ \bibnamefont
  {Sultanov}}, \bibinfo {author} {\bibfnamefont {Maximilian~A.}\ \bibnamefont
  {Weissflog}}, \bibinfo {author} {\bibfnamefont {Dennis}\ \bibnamefont
  {Arslan}}, \bibinfo {author} {\bibfnamefont {Mohammadreza}\ \bibnamefont
  {Younesi}}, \bibinfo {author} {\bibfnamefont {Thomas}\ \bibnamefont
  {Pertsch}}, \bibinfo {author} {\bibfnamefont {Isabelle}\ \bibnamefont
  {Staude}}, \bibinfo {author} {\bibfnamefont {Frank}\ \bibnamefont
  {Setzpfandt}}, \ and\ \bibinfo {author} {\bibfnamefont {Maria~V.}\
  \bibnamefont {Chekhova}},\ }\href@noop {} {\enquote {\bibinfo {title}
  {Spontaneous parametric down-conversion from resonant metasurfaces},}\ }
  (\bibinfo {year} {2021}),\ \Eprint {http://arxiv.org/abs/2103.08524}
  {arXiv:2103.08524 [physics.optics]} \BibitemShut {NoStop}%
\bibitem [{\citenamefont {Sinev}\ \emph {et~al.}(2017)\citenamefont {Sinev},
  \citenamefont {Bogdanov}, \citenamefont {Komissarenko}, \citenamefont
  {Frizyuk}, \citenamefont {Petrov}, \citenamefont {Mukhin}, \citenamefont
  {Makarov}, \citenamefont {Samusev}, \citenamefont {Lavrinenko},\ and\
  \citenamefont {Iorsh}}]{2017_Sinev}%
  \BibitemOpen
  \bibfield  {author} {\bibinfo {author} {\bibfnamefont {Ivan~S.}\ \bibnamefont
  {Sinev}}, \bibinfo {author} {\bibfnamefont {Andrey~A.}\ \bibnamefont
  {Bogdanov}}, \bibinfo {author} {\bibfnamefont {Filipp~E.}\ \bibnamefont
  {Komissarenko}}, \bibinfo {author} {\bibfnamefont {Kristina~S.}\ \bibnamefont
  {Frizyuk}}, \bibinfo {author} {\bibfnamefont {Mihail~I.}\ \bibnamefont
  {Petrov}}, \bibinfo {author} {\bibfnamefont {Ivan~S.}\ \bibnamefont
  {Mukhin}}, \bibinfo {author} {\bibfnamefont {Sergey~V.}\ \bibnamefont
  {Makarov}}, \bibinfo {author} {\bibfnamefont {Anton~K.}\ \bibnamefont
  {Samusev}}, \bibinfo {author} {\bibfnamefont {Andrei~V.}\ \bibnamefont
  {Lavrinenko}}, \ and\ \bibinfo {author} {\bibfnamefont {Ivan~V.}\
  \bibnamefont {Iorsh}},\ }\bibfield  {title} {\enquote {\bibinfo {title}
  {Chirality driven by magnetic dipole response for demultiplexing of surface
  waves},}\ }\href {\doibase 10.1002/lpor.201700168} {\bibfield  {journal}
  {\bibinfo  {journal} {Laser {\&} Photonics Reviews}\ }\textbf {\bibinfo
  {volume} {11}},\ \bibinfo {pages} {1700168} (\bibinfo {year}
  {2017})}\BibitemShut {NoStop}%
\bibitem [{\citenamefont {Sinev}\ \emph {et~al.}(2020)\citenamefont {Sinev},
  \citenamefont {Komissarenko}, \citenamefont {Iorsh}, \citenamefont
  {Permyakov}, \citenamefont {Samusev},\ and\ \citenamefont
  {Bogdanov}}]{2020_Sinev}%
  \BibitemOpen
  \bibfield  {author} {\bibinfo {author} {\bibfnamefont {Ivan}\ \bibnamefont
  {Sinev}}, \bibinfo {author} {\bibfnamefont {Filipp}\ \bibnamefont
  {Komissarenko}}, \bibinfo {author} {\bibfnamefont {Ivan}\ \bibnamefont
  {Iorsh}}, \bibinfo {author} {\bibfnamefont {Dmitry}\ \bibnamefont
  {Permyakov}}, \bibinfo {author} {\bibfnamefont {Anton}\ \bibnamefont
  {Samusev}}, \ and\ \bibinfo {author} {\bibfnamefont {Andrey}\ \bibnamefont
  {Bogdanov}},\ }\bibfield  {title} {\enquote {\bibinfo {title} {Steering of
  guided light with dielectric nanoantennas},}\ }\href {\doibase
  10.1021/acsphotonics.9b01515} {\bibfield  {journal} {\bibinfo  {journal}
  {{ACS} Photonics}\ }\textbf {\bibinfo {volume} {7}},\ \bibinfo {pages}
  {680--686} (\bibinfo {year} {2020})}\BibitemShut {NoStop}%
\bibitem [{\citenamefont {Poddubny}\ \emph {et~al.}(2016)\citenamefont
  {Poddubny}, \citenamefont {Iorsh},\ and\ \citenamefont
  {Sukhorukov}}]{2016_Poddubny}%
  \BibitemOpen
  \bibfield  {author} {\bibinfo {author} {\bibfnamefont {Alexander~N.}\
  \bibnamefont {Poddubny}}, \bibinfo {author} {\bibfnamefont {Ivan~V.}\
  \bibnamefont {Iorsh}}, \ and\ \bibinfo {author} {\bibfnamefont {Andrey~A.}\
  \bibnamefont {Sukhorukov}},\ }\bibfield  {title} {\enquote {\bibinfo {title}
  {Generation of photon-plasmon quantum states in nonlinear hyperbolic
  metamaterials},}\ }\href {\doibase 10.1103/physrevlett.117.123901} {\bibfield
   {journal} {\bibinfo  {journal} {Physical Review Letters}\ }\textbf {\bibinfo
  {volume} {117}},\ \bibinfo {pages} {123901} (\bibinfo {year}
  {2016})}\BibitemShut {NoStop}%
\bibitem [{\citenamefont {Lenzini}\ \emph {et~al.}(2018)\citenamefont
  {Lenzini}, \citenamefont {Poddubny}, \citenamefont {Titchener}, \citenamefont
  {Fisher}, \citenamefont {Boes}, \citenamefont {Kasture}, \citenamefont
  {Haylock}, \citenamefont {Villa}, \citenamefont {Mitchell}, \citenamefont
  {Solntsev}, \citenamefont {Sukhorukov},\ and\ \citenamefont
  {Lobino}}]{2018_Lenzini}%
  \BibitemOpen
  \bibfield  {author} {\bibinfo {author} {\bibfnamefont {Francesco}\
  \bibnamefont {Lenzini}}, \bibinfo {author} {\bibfnamefont {Alexander~N}\
  \bibnamefont {Poddubny}}, \bibinfo {author} {\bibfnamefont {James}\
  \bibnamefont {Titchener}}, \bibinfo {author} {\bibfnamefont {Paul}\
  \bibnamefont {Fisher}}, \bibinfo {author} {\bibfnamefont {Andreas}\
  \bibnamefont {Boes}}, \bibinfo {author} {\bibfnamefont {Sachin}\ \bibnamefont
  {Kasture}}, \bibinfo {author} {\bibfnamefont {Ben}\ \bibnamefont {Haylock}},
  \bibinfo {author} {\bibfnamefont {Matteo}\ \bibnamefont {Villa}}, \bibinfo
  {author} {\bibfnamefont {Arnan}\ \bibnamefont {Mitchell}}, \bibinfo {author}
  {\bibfnamefont {Alexander~S}\ \bibnamefont {Solntsev}}, \bibinfo {author}
  {\bibfnamefont {Andrey~A}\ \bibnamefont {Sukhorukov}}, \ and\ \bibinfo
  {author} {\bibfnamefont {Mirko}\ \bibnamefont {Lobino}},\ }\bibfield  {title}
  {\enquote {\bibinfo {title} {Direct characterization of a nonlinear photonic
  circuit's wave function with laser light},}\ }\href {\doibase
  10.1038/lsa.2017.143} {\bibfield  {journal} {\bibinfo  {journal} {Light:
  Science {\&} Applications}\ }\textbf {\bibinfo {volume} {7}},\ \bibinfo
  {pages} {17143--17143} (\bibinfo {year} {2018})}\BibitemShut {NoStop}%
\bibitem [{\citenamefont {Poddubny}\ and\ \citenamefont
  {Smirnova}(2018)}]{2018_Poddubny}%
  \BibitemOpen
  \bibfield  {author} {\bibinfo {author} {\bibfnamefont {Alexander~N.}\
  \bibnamefont {Poddubny}}\ and\ \bibinfo {author} {\bibfnamefont {Daria~A.}\
  \bibnamefont {Smirnova}},\ }\href@noop {} {\enquote {\bibinfo {title}
  {Nonlinear generation of quantum-entangled photons from high-q states in
  dielectric nanoparticles},}\ } (\bibinfo {year} {2018}),\ \Eprint
  {http://arxiv.org/abs/1808.04811} {arXiv:1808.04811 [physics.optics]}
  \BibitemShut {NoStop}%
\bibitem [{\citenamefont {Novotny}\ and\ \citenamefont
  {Hecht}(2012)}]{Novotny_2012}%
  \BibitemOpen
  \bibfield  {author} {\bibinfo {author} {\bibfnamefont {Lukas}\ \bibnamefont
  {Novotny}}\ and\ \bibinfo {author} {\bibfnamefont {Bert}\ \bibnamefont
  {Hecht}},\ }\href
  {http://www.amazon.com/Principles-Nano-Optics-Lukas-Novotny/dp/1107005469}
  {\emph {\bibinfo {title} {{Principles of Nano-Optics}}}}\ (\bibinfo
  {publisher} {Cambridge University Press},\ \bibinfo {year} {2012})\ p.\
  \bibinfo {pages} {578}\BibitemShut {NoStop}%
\bibitem [{\citenamefont {Kostina}\ \emph {et~al.}(2019)\citenamefont
  {Kostina}, \citenamefont {Petrov}, \citenamefont {Ivinskaya}, \citenamefont
  {Sukhov}, \citenamefont {Bogdanov}, \citenamefont {Toftul}, \citenamefont
  {Nieto-Vesperinas}, \citenamefont {Ginzburg},\ and\ \citenamefont
  {Shalin}}]{2019_Kostina}%
  \BibitemOpen
  \bibfield  {author} {\bibinfo {author} {\bibfnamefont {Natalia}\ \bibnamefont
  {Kostina}}, \bibinfo {author} {\bibfnamefont {Mihail}\ \bibnamefont
  {Petrov}}, \bibinfo {author} {\bibfnamefont {Aliaksandra}\ \bibnamefont
  {Ivinskaya}}, \bibinfo {author} {\bibfnamefont {Sergey}\ \bibnamefont
  {Sukhov}}, \bibinfo {author} {\bibfnamefont {Andrey}\ \bibnamefont
  {Bogdanov}}, \bibinfo {author} {\bibfnamefont {Ivan}\ \bibnamefont {Toftul}},
  \bibinfo {author} {\bibfnamefont {Manuel}\ \bibnamefont {Nieto-Vesperinas}},
  \bibinfo {author} {\bibfnamefont {Pavel}\ \bibnamefont {Ginzburg}}, \ and\
  \bibinfo {author} {\bibfnamefont {Alexander}\ \bibnamefont {Shalin}},\
  }\bibfield  {title} {\enquote {\bibinfo {title} {Optical binding via surface
  plasmon polariton interference},}\ }\href {\doibase
  10.1103/physrevb.99.125416} {\bibfield  {journal} {\bibinfo  {journal}
  {Physical Review B}\ }\textbf {\bibinfo {volume} {99}},\ \bibinfo {pages}
  {125416} (\bibinfo {year} {2019})}\BibitemShut {NoStop}%
\bibitem [{\citenamefont {Johnson}\ and\ \citenamefont
  {Christy}(1972)}]{1972_Johnson}%
  \BibitemOpen
  \bibfield  {author} {\bibinfo {author} {\bibfnamefont {P.~B.}\ \bibnamefont
  {Johnson}}\ and\ \bibinfo {author} {\bibfnamefont {R.~W.}\ \bibnamefont
  {Christy}},\ }\bibfield  {title} {\enquote {\bibinfo {title} {Optical
  constants of the noble metals},}\ }\href {\doibase 10.1103/physrevb.6.4370}
  {\bibfield  {journal} {\bibinfo  {journal} {Physical Review B}\ }\textbf
  {\bibinfo {volume} {6}},\ \bibinfo {pages} {4370--4379} (\bibinfo {year}
  {1972})}\BibitemShut {NoStop}%
\bibitem [{\citenamefont {Boyd}(2008)}]{Boyd_2008}%
  \BibitemOpen
  \bibfield  {author} {\bibinfo {author} {\bibfnamefont {Robert}\ \bibnamefont
  {Boyd}},\ }\href
  {https://www.elsevier.com/books/nonlinear-optics/boyd/978-0-12-369470-6}
  {\emph {\bibinfo {title} {{Nonlinear Optics (3rd ed.)}}}}\ (\bibinfo
  {publisher} {Academic Press},\ \bibinfo {year} {2008})\ p.\ \bibinfo {pages}
  {640}\BibitemShut {NoStop}%
\bibitem [{\citenamefont {Ekert}\ and\ \citenamefont
  {Knight}(1995)}]{1995_Ekert}%
  \BibitemOpen
  \bibfield  {author} {\bibinfo {author} {\bibfnamefont {Artur}\ \bibnamefont
  {Ekert}}\ and\ \bibinfo {author} {\bibfnamefont {Peter~L.}\ \bibnamefont
  {Knight}},\ }\bibfield  {title} {\enquote {\bibinfo {title} {Entangled
  quantum systems and the {Schmidt} decomposition},}\ }\href {\doibase
  10.1119/1.17904} {\bibfield  {journal} {\bibinfo  {journal} {American Journal
  of Physics}\ }\textbf {\bibinfo {volume} {63}},\ \bibinfo {pages} {415--423}
  (\bibinfo {year} {1995})}\BibitemShut {NoStop}%
\bibitem [{\citenamefont {Law}\ and\ \citenamefont {Eberly}(2004)}]{2004_Law}%
  \BibitemOpen
  \bibfield  {author} {\bibinfo {author} {\bibfnamefont {C.~K.}\ \bibnamefont
  {Law}}\ and\ \bibinfo {author} {\bibfnamefont {J.~H.}\ \bibnamefont
  {Eberly}},\ }\bibfield  {title} {\enquote {\bibinfo {title} {Analysis and
  interpretation of high transverse entanglement in optical parametric down
  conversion},}\ }\href {\doibase 10.1103/physrevlett.92.127903} {\bibfield
  {journal} {\bibinfo  {journal} {Physical Review Letters}\ }\textbf {\bibinfo
  {volume} {92}},\ \bibinfo {pages} {127903} (\bibinfo {year}
  {2004})}\BibitemShut {NoStop}%
\bibitem [{\citenamefont {Ghirardi}\ and\ \citenamefont
  {Marinatto}(2004)}]{2004_Ghirardi}%
  \BibitemOpen
  \bibfield  {author} {\bibinfo {author} {\bibfnamefont {GianCarlo}\
  \bibnamefont {Ghirardi}}\ and\ \bibinfo {author} {\bibfnamefont {Luca}\
  \bibnamefont {Marinatto}},\ }\bibfield  {title} {\enquote {\bibinfo {title}
  {General criterion for the entanglement of two indistinguishable
  particles},}\ }\href {\doibase 10.1103/physreva.70.012109} {\bibfield
  {journal} {\bibinfo  {journal} {Physical Review A}\ }\textbf {\bibinfo
  {volume} {70}},\ \bibinfo {pages} {012109} (\bibinfo {year}
  {2004})}\BibitemShut {NoStop}%
\end{thebibliography}

\end{document}